\documentclass[twocolumn]{aastex631}

\usepackage{xspace}
\usepackage{amsmath}
\usepackage{multirow}
\usepackage[T1]{fontenc}
\usepackage{gensymb}

\newcommand{\spitzer}{{\textit{Spitzer}}\xspace}
\newcommand{\hst}{{\textit{HST}}\xspace}
\newcommand{\jwst}{{\textit{JWST}}\xspace}

\newcommand{\planetname}{{LTT~1445Ab}\xspace}

\newcommand{\uviswvl}{$\rm 0.2-0.8\;\mu m$}
\newcommand{\irwvl}{$\rm 1.1-1.65\;\mu m$}

\newcommand{\eureka}{\texttt{Eureka!}\xspace}
\newcommand{\firefly}{\texttt{FIREFLy}\xspace}
\newcommand{\exotic}{\texttt{ExoTiC-ISM}\xspace}

\newcommand{\POSEIDON}{\texttt{POSEIDON}\xspace}

\newcommand{\magenta}[1]{\textcolor{magenta}{#1}}

\defcitealias{Lustig-Yaeger2023}{Lustig-Yaeger \& Fu et al.}
\defcitealias{Moran2023}{Moran \& Stevenson et al.}
\defcitealias{May2023}{May \& MacDonald et al.}

\begin{document}

\title{An \textit{HST} Transmission Spectrum of the Closest M-Dwarf Transiting Rocky Planet LTT~1445Ab}

\author[0000-0002-9030-0132]{Katherine A. Bennett}
\affiliation{Department of Earth \& Planetary Sciences, Johns Hopkins University, Baltimore, MD 21218, USA}

\author[0000-0001-6050-7645]{David K. Sing}
\affiliation{Department of Earth \& Planetary Sciences, Johns Hopkins University, Baltimore, MD 21218, USA}
\affiliation{Department of Physics \& Astronomy, Johns Hopkins University, Baltimore, MD, 21218 USA}

\author[0000-0002-7352-7941]{Kevin B. Stevenson}
\affiliation{Johns Hopkins APL, Laurel, MD 20723, USA}
\affil{Consortium on Habitability and Atmospheres of M-dwarf Planets (CHAMPs), Laurel, MD, USA}

\author[0000-0003-4328-3867]{Hannah R. Wakeford}
\affil{School of Physics, HH Wills Physics Laboratory, University of Bristol, Bristol, BS8 1TL, UK}
\affil{Consortium on Habitability and Atmospheres of M-dwarf Planets (CHAMPs), Laurel, MD, USA}

\author[0000-0003-4408-0463]{Zafar Rustamkulov}
\affiliation{Department of Earth \& Planetary Sciences, Johns Hopkins University, Baltimore, MD 21218, USA}

\author[0000-0002-0832-710X]{Natalie H. Allen}
\affiliation{Department of Physics \& Astronomy, Johns Hopkins University, Baltimore, MD, 21218 USA}

\author[0000-0003-3667-8633]{Joshua D. Lothringer}
\affiliation{Space Telescope Science Institute, Baltimore, MD 21218, USA}

\author[0000-0003-4816-3469]{Ryan J. MacDonald}
\affiliation{Department of Astronomy, University of Michigan, 1085 S. University Ave., Ann Arbor, MI 48109, USA}
\affiliation{NHFP Sagan Fellow}

\author[0000-0001-6707-4563]{Nathan J. Mayne}
\affil{Department of Physics \& Astronomy, University of Exeter, Exeter, EX4 4QF, UK}

\author[0000-0002-3263-2251]{Guangwei Fu}
\affiliation{Department of Physics \& Astronomy, Johns Hopkins University, Baltimore, MD, 21218 USA}

\begin{abstract}

Which rocky exoplanets have atmospheres? This presumably simply question is the first that must be answered to understand the prevalence of nearby habitable planets. A mere 6.9 pc from Earth, LTT~1445A is the closest transiting M-dwarf system, and its largest known planet, at $\rm 1.31\; R_{\oplus}$ and 424~K, is one of the most promising targets in which to search for an atmosphere. We use \textit{HST}/WFC3 transmission spectroscopy with the G280 and G141 grisms to study the spectrum of LTT~1445Ab between $\rm 0.2-1.65\;\mu m$. In doing so, we uncover a UV flare on the neighboring star LTT~1445C that is completely invisible at optical wavelengths; we report one of the first simultaneous near-UV/optical spectra of an M~dwarf flare. The planet spectrum is consistent with a flat line (with median transit depth uncertainties of 128 and 52 ppm for the G280 and G141 observations, respectively), though the infrared portion displays potential features that could be explained by known opacity sources such as HCN. Some atmospheric retrievals weakly favor ($\sim2\sigma$) an atmosphere, but it remains challenging to discern between stellar contamination, an atmosphere, and a featureless spectrum at this time. We do, however, confidently rule out $\leq100\times$ solar metallicity atmospheres. Although stellar contamination retrievals cannot fit the infrared features well, the overall spectrum is consistent with stellar contamination from hot or cold spots. Based on the UV/optical data, we place limits on the extent of stellar variability expected in the near-infrared ($30-40$ ppm), which will be critical for future \textit{JWST} observations. 

\end{abstract}


\section{Introduction} \label{sec:intro}
 
With the explosion of small planet detections in the last decade, 
we are now in the first stages of uncovering the preponderance of habitable planets in our Galaxy. While there are many avenues of explanations for what makes a planet ``habitable", the most observationally accessible is the simple question of whether or not an atmosphere is present on a rocky exoplanet.

This question is most readily answerable for rocky planets around M dwarfs, as our observing strategies are optimal for the large planet-to-star mass and radius ratios found around these smaller stars. Additionally, because M dwarfs are the coolest stars, the habitable zones of these systems are much closer in to the host stars. Together, this means we can observe multiple transits in quick succession of small, possibly habitable rocky planets that otherwise would be unobservable around larger stars. Coined the ``M-dwarf opportunity" (e.g., \citealt{Charbonnaeu2007}), this strategy has largely sculpted the scientific focus of the rocky exoplanet community. 

Naturally, however, there is a catch. While they are the most optimal observationally, M~dwarfs also emit high amounts of X-ray and extreme-ultraviolet (XUV) radiation over a prolonged period of time as they contract onto the main sequence (e.g., \citealt{Preibisch2005, West2008, Peacock2020}), as well as display higher rates of stellar flares and/or coronal mass ejections (e.g., \citealt{Walker1981, Audard2000, Youngblood2017, Howard2023}). Both of these factors may drive partial or complete atmospheric erosion of rocky planets early on in their geologic history (e.g., \citealt{Khodachenko2007, Tian2009, Owen2012, Peacock2019, Kite2020, vanLooveren2024}). Thus, understanding which M-dwarf rocky planets are able to retain their atmospheres against these extreme conditions is the most critical and outstanding question in rocky exoplanet astronomy. 

A planet's ability to retain its atmosphere is likely due to some combination of planet size and stellar insolation, where larger, cooler planets are more easily able to retain atmospheres. More massive planets have a higher escape velocity ($v_{\rm esc}$), making it more difficult for atmospheric gases to escape, while planets experiencing lower stellar insolation ($I$) are less likely to have their atmospheres undergo mass loss. This mass loss can be driven by several escape processes, including thermal escape, in which gases are heated in the exosphere such that their thermal velocities exceed their escape velocities, and hydrodynamic escape, in which the entirety of the upper atmosphere acts as a fluid and is lost to space \citep{Owen2019}. 

The dividing line between planets with and without atmospheres has been coined the ``cosmic shoreline" \citep{Zahnle2017}. In the Solar System, this dividing line goes as $I\propto v_{\rm esc}^4$. However, in exoplanets, the dependence on these parameters is highly unconstrained. While the cosmic shoreline is widely hypothesized to be a ubiquitous phenomenon, driving exoplanet demographics in the Galaxy, the importance of thermal versus hydrodynamic escape and the precise role of stellar XUV insolation remains to be elucidated. The truth is that we simply do not know yet whether this paradigm can be used as a tool to predict which of our terrestrial neighbors have atmospheres. 

It is therefore critical to determine whether or not individual rocky planets have atmospheres in order to 1) search for possibly habitable planets in the nearby Galaxy, and 2) understand rocky exoplanets on a population level and determine whether all stellar systems share a common cosmic shoreline. 

\subsection{The Ongoing Importance of \hst}

While the \textit{James Webb Space Telescope} (\jwst) has been critical in our advancement of the search for atmospheres on rocky exoplanets, the \textit{Hubble Space Telescope} (\hst) still has a very useful role to play, particularly with bright sources such as LTT~1445A. 
Specifically, the Wide Field Camera 3 (WFC3) instrument provides spectroscopic information in the ultraviolet-visible (UVIS) and near-infrared (NIR) with different grisms, which can be used to probe stellar contamination and place lower limits on the mean molecular weight of any atmosphere present.

\hst/WFC3's G141 observing mode probes the $\rm 1.1-1.65\;\mu m$ region. This includes the $\rm 1.4\;\mu m$ $\rm H_2O$ feature, which, pre-\jwst, has been one of the most widely detected features in exoplanet atmospheres (e.g., \citealt{Deming2013, Wakeford2013, Fraine2014, Kreidberg2014b, McCullough2014, Damiano2017, Changeat2020, Roy2023}). While unable to reach the precision afforded by \jwst, \hst's WFC3/G141 mode has been widely used to place constraints on the presence of primordial or hydrogen-dominated atmospheres in rocky exoplanets, as these low mean molecular weight atmospheres would lead to large features in the transmission spectrum that are measurable by WFC3/G141. So far, no cloud-free $\rm H_2/He$-rich atmospheres around rocky M-dwarf planets have been detected with \hst: TRAPPIST-1b and c \citep{deWit2016, Zhang2018}; TRAPPIST-1d, e, and f \citep{deWit2018, Zhang2018}; L98-59b \citep{Damiano2022, Zhou2022}; L98-59c and d \citep{Zhou2023}; and GJ~1132b \citep{Mugnai2021, Libby-Roberts2022}.

Additionally, and perhaps more critically, \hst's WFC3/G280 grism (covering approximately $\rm 0.2-0.8\;\mu m$) provides the most efficient method to probe exoplanet atmospheres in the near-UV (NUV) and visible wavelengths. While a combination of gratings with \hst's Space Telescope Imaging Spectrograph (STIS) instrument can also achieve comparable wavelength coverage, this requires multiple observations with multiple gratings. The efficient, broad wavelength coverage achievable with UVIS/G280 opens up the use of this observing mode for exoplanet transmission spectroscopy, as recently demonstrated by \cite{Wakeford2020} and \cite{Lothringer2022}. 

Indeed, our study marks the first time the UVIS/G280 mode is employed for transmission spectroscopy of a rocky exoplanet. For the case of smaller planets, the UV-optical wavelengths are still critical, in addition to the longer IR wavelengths afforded to us by \jwst. This is because the UVIS wavelengths can reveal the presence of stellar contamination through a steep slope in the UV-optical (e.g., \citealt{Pont2008, Sing2011}), which occurs because of the stronger contrast between spot and photospheric temperatures at shorter wavelengths. The importance of UV spectroscopy to exoplanet astronomy was recently demonstrated in the report by the working group on the Strategic Exoplanet Initiatives with \hst and \jwst \citep{Redfield2024}, which was adopted by the director of the Space Telescope Science Institute (STScI) in July 2024. The report recommends mid-infrared emission \jwst studies of nearby rocky M-dwarf exoplanets alongside near-simultaneous UV studies with \hst in order to fully characterize the system's UV environment.

\subsection{The LTT~1445 System}

\planetname (alias: TOI-455b; \citealt{Winters2019, Oddo2023}) is our nearest M-dwarf transiting rocky neighbor and provides one of the best chances at detecting an atmosphere around a rocky planet. At 424 K (assuming $A_b=0$ and complete heat redistribution) and $\rm 1.31\;R_{\oplus}$ \citep{Winters2022}, it is one of the coolest and largest rocky planets discovered to date, and it should provide a cogent test of the cosmic shoreline hypothesis.

Recently discovered in 2019 \citep{Winters2019}, it is part of a hierarchical triplet system of M dwarfs, which in and of itself provides an intriguing laboratory for M-dwarf stellar astrophysics. The system parameters are shown in \autoref{tab:sys_params}. \planetname orbits the primary star, LTT~1445A, which is separated from the (often blended) B and C stars by approximately 34 au \citep{Winters2019}, on average. The orbit of the A-BC system is estimated to be about 250 years \citep{Winters2019}. LTT~1445B and C are separated by 8.1 au and orbit each other every $36.2\pm5.3$ years \citep{Winters2019}. See the bottom right panel of \autoref{fig:raw_data} for an \hst acquisition image of the triplet system. We note that, in 2003,  LTT~1445B was the star farthest from LTT~1445A (see \citealt{Winters2019}, their Figure 1, for an archival \hst/NICMOS image), but by 2023 LTT~1445C was instead the farthest, as established through our flux measurements (C is the smallest and emits the lowest flux). Interestingly, the fact that all three stars continue to fall on a line, though B and C have reversed position relative to A, suggests that the entire system is indeed coplanar, as suggested by \cite{Winters2022}. 

From our acquisition images, we further corroborate the finding from \cite{Winters2019} that the separation between the A and BC components is decreasing. From the 2021 \hst/WFC3 G141 acquision image (shown on the bottom right of \autoref{fig:raw_data}) to the 2023 \hst/WFC3 G280 acquisition image (not pictured), we calculate that the sky-projected angular separation between A and B decreased from 7.47" to 6.74" (corresponding to a decrease from 51 to 46 au, assuming a parallax of 145.6922 mas, \citealt{gaiadr3}). Meanwhile, the sky-projected angular separation between B and C increased from 1.28" to 1.37" (corresponding to sky-projected distances of 8.8 to 9.4 au) over nearly three years.

\begin{deluxetable*}{ccccc}
\tablecaption{Parameters of the LTT~1445 system. \label{tab:sys_params}}
\tablenum{1}
\tablewidth{0pt}
\def\arraystretch{.85}
\tablehead{\colhead{Parameter} & \multicolumn{2}{c}{Value}}
\startdata
         & \multicolumn{2}{c}{\textbf{LTT~1445A}} &  \textbf{LTT~1445B} & \textbf{LTT~1445C} \\
         Distance (pc) & \multicolumn{2}{c}{$6.869\pm0.004^{\rm a}$} & $6.869\pm0.004^{\rm a}$ & $6.869\pm0.004^{\rm a}$ \\
         $V_J$ magnitude & \multicolumn{2}{c}{$11.22\pm0.02^{\rm b}$} & $11.78\pm0.09^{\rm c}$ & $12.64\pm0.09^{\rm c}$\\
         Stellar Radius ($\rm R_{\odot}$) & \multicolumn{2}{c}{$0.265^{+0.011}_{-0.010}$} & $0.236\pm0.027^{\rm c}$ & $0.197\pm0.027^{\rm c}$ \\ 
         Stellar Mass ($\rm M_{\odot}$) & \multicolumn{2}{c}{$0.257\pm0.014$} & $0.215\pm0.014^{\rm c}$ & $0.161\pm0.014^{\rm c}$ \\
         Spectral Type & \multicolumn{2}{c}{$\rm M3.5V^{\rm d}$} & $\rm M3.5V^{\rm d}$ & $\rm M4V^{\rm d}$\\
         Stellar $T_{\rm eff}\,$ (K) & \multicolumn{2}{c}{$3340\pm150$} & $-$ & $-$\\
         log $g_{\odot}$ $(\rm cm\;s^{-2}$) & \multicolumn{2}{c}{$4.967^{+0.061\; {\rm c}}_{-0.075}$} & $-$ & $-$ \\
         $\rm [Fe/H]$ & \multicolumn{2}{c}{$-0.34\pm0.09$} & $-$ & $-$\\
         $P_{\rm rot}$ (days) & \multicolumn{2}{c}{$85\pm22$} & $-$ & $-$\\
         log $\rm R^{'}_{HK}$ & \multicolumn{2}{c}{$-5.413\pm0.118$} & $-$ & $-$\\
         Age (Gyr) & \multicolumn{2}{c}{\magenta{$>2.2^{\rm e}$}} & $-$ & $-$\\
         \hline
         & \textbf{\planetname} & \textbf{LTT~1445Ac} & & \\
         Planet Radius ($\rm R_{\oplus}$) & $1.305^{+0.066}_{-0.061}$ & $1.147^{+0.055}_{-0.054}$\\ 
         Planet Mass ($\rm M_{\oplus}$) & $2.87^{+0.26}_{-0.25}$ & $1.54^{+0.20}_{-0.19}$\\
         Planet Density ($\rm g\;cm^{-3}$) & $7.1^{+1.2}_{-1.1}$ & $5.57^{+0.68}_{-0.60}$\\
         log $g_P$ $(\rm cm\;s^{-2}$) & $3.217^{+0.050}_{-0.053}$ & $3.057^{+0.042}_{-0.043}$\\
         Planetary $T_{\rm eq}$ (K) & $424\pm21$ & $508\pm25$ \\
         $a$ (au) & $0.03813^{+0.00068}_{-0.00070}$ & $0.02661^{+0.00047}_{-0.00049}$\\
         $P$ (days) & $5.3587657^{+0.0000043}_{-0.0000042}$ & $3.1239035^{+0.0000034}_{-0.0000036}$ \\
         Inclination (degrees) & $89.68^{+0.22}_{-0.29}$ & $87.43^{+0.18}_{-0.29}$\\
         Transit Duration (min) & $82.04^{+1.02}_{-0.98}$ & $28.9\pm1.6$ \\
         $T_{0}\;(\rm BJD_{TDB})$ & $2458412.70851^{+0.00040}_{-0.00039}$ & $2458412.58156^{+0.00059}_{-0.00057}$\\
\enddata
\tablecomments{Values taken from \cite{Winters2022} unless otherwise noted. a. \cite{Lindegren2021}; b. \cite{Henry2018}; c. \cite{Winters2019}; d. \cite{Brown2022}; e. \cite{Diamond-Lowe2024}. $T_{\rm eq}$ is calculated assuming $A_B=0$ and perfect redistribution.}
\end{deluxetable*}

\planetname's radius ($\rm 1.305^{+0.066}_{-0.061}\;R_{\oplus}$) and mass ($\rm 2.87^{+0.26}_{-0.25}\;M_{\oplus}$) give it a density of $7.1^{+1.2}_{-1.1}\;\rm g\;cm^{-3}$, consistent with a rocky nature (see Figure 10 in \citealt{Winters2022}). We note that \cite{Oddo2023} derived a slightly lower planetary radius of $\rm 1.18\pm0.06\;R_{\oplus}$, but the authors recognize that their calculated radius may be discrepant from that in both \cite{Winters2019} and \cite{Winters2022} (which agree to within 1.3$\sigma$) because of the difficulties in disentangling the three stellar sources. 

\planetname orbits its star every 5.36 days, and it is accompanied by another transiting sibling, LTT~1445Ac, which has a period of 3.12 days \citep{Winters2022, Lavie2023}. A third non-transiting planet candidate was detected by \citep{Lavie2023}. Like its larger sibling, the mass and radius of the hotter and smaller LTT~1445Ac are consistent with a rocky nature \citep{Pass2023}. As one of the few known multi-planet M-dwarf systems, LTT~1445A offers a unique opportunity for comparative exoplanetology. 

\planetname has recently been observed via low-resolution ground-based transmission spectroscopy. \cite{Diamond-Lowe2023} observed the system from $0.6-1\rm\;\mu m$ using Magellan II/LDSS3C and ruled out a cloud-free, $1\times$ solar metallicity atmosphere for 1 and 10 bars of surface pressure. 

In this paper, we present the first space-based observations of \planetname using \hst/WFC3's UVIS (G280) and IR (G141) grisms. These observations can be used for several purposes, including 1) as a method to assess the level of stellar contamination present and 2) as a first-step reconnaissance of the planet's IR transmission spectrum. \jwst has observed \planetname in emission with the Mid-Infrared Instrument (MIRI; GO Program 2708; PI Z. Berta-Thompson), and it will also observe the planet in transmission with the Near-Infrared Spectrograph (NIRSpec) later this year (GO Program 2512; PI N. Batalha). Taken together, these transmission observations will provide some of the broadest wavelength coverage of a rocky planet transmission spectrum to date, while any emission from the eclipse observation should corroborate the transmission findings.

Our study is structured as follows. Section \ref{sec:obs} describes the observations before Section \ref{sec:reduction} details our data reduction process for both the UVIS and IR data. We present our spectral interpretation in Section \ref{sec:analysis} and, finally, discuss the implications of our findings in Section \ref{sec:discussion}.

\section{Observations} \label{sec:obs}

We obtained four transits of \planetname between 2020 and 2023 with \hst/WFC3 (Wide Field Camera 3) as part of GO Program 16039 (PI D. Sing). Three of those transits used the UVIS (ultraviolet-visible) channel CCD detector with the G280 grism, which covers the wavelength range \uviswvl. The G280 observing mode has recently been implemented for exoplanet time series observations, and to date, this is only the third planet to be observed in this mode \citep{Wakeford2020, Lothringer2022}. See \cite{Wakeford2020} for details about the G280 grism, as well as its challenges and advantages for the exoplanet community. 

One transit used the infrared (IR) channel detector with the G141 grism, covering \irwvl. The program intended for a G102 transit as well, but the observations failed and proved unschedulable, so an additional (third) UVIS transit was observed instead. The UVIS transits occurred on 2020~September~27, 2020~December~11, and 2023~October~11, and the IR transit occurred on 2021~January~07. 

The first two UVIS transit observations span three \hst orbits, while the third span four orbits. The UVIS CCD contains two chips, and we utilize a $2250\times590$ subarray centered on the second chip, which has been shown to be the more stable of the two \citep{Wakeford2020}. To do so, we use a 50" offset in the y-position target reference system, with the subarray readout centered on pixels 2136 (CENTERAXIS1) and 1216 (CENTERAXIS2). Our requested orient ranges were 38--148\degree\;or 218--328\degree. 
At the beginning of each observation, we take an aquisition image to perform wavelength calibration, followed by 9--10 exposures per orbit (with each exposure lasting 195 seconds for UVIS visits 1 and 2 and 185.5 seconds for visit 3) for a total of 29 exposures for the first two visits and 39 exposures for the third visit.

The G141 (IR) observations utilize an HgCdTe detector, which performs an ``up-the-ramp" sampling technique to measure the electron count rate, similar to the method employed by \jwst. For the IR transit, we use six groups per integration and 190 total integrations over four orbits. We use the spatial scanning mode for the IR observations, in which the telescope moves slightly during the observation, creating forward and reverse scans, in order to spread the light in the cross-dispersion direction. Useful for bright stars ($V_{mag}<11$; \citealt{Wakeford2016}) and implemented in \hst Cycle 19, this technique greatly boosts signal-to-noise (S/N) and reduces risk of saturation while decreasing overheads \citep{Wakeford2013}. It is now the dominant observing mode for WFC3 IR transit observations (e.g., \citealt{Carone2021, Garcia2022, Libby-Roberts2022}) and should be used whenever the target is too bright for stare mode. In order to ensure the spectra for stars B and C appear completely on the detector, we apply a 7" offset in the y-position target reference system. For WFC3/G141, our requested orient ranges were 53--126\degree\;or 233--306\degree.

\section{Data Reduction} \label{sec:reduction}

\begin{figure*}
    \centering
    \includegraphics{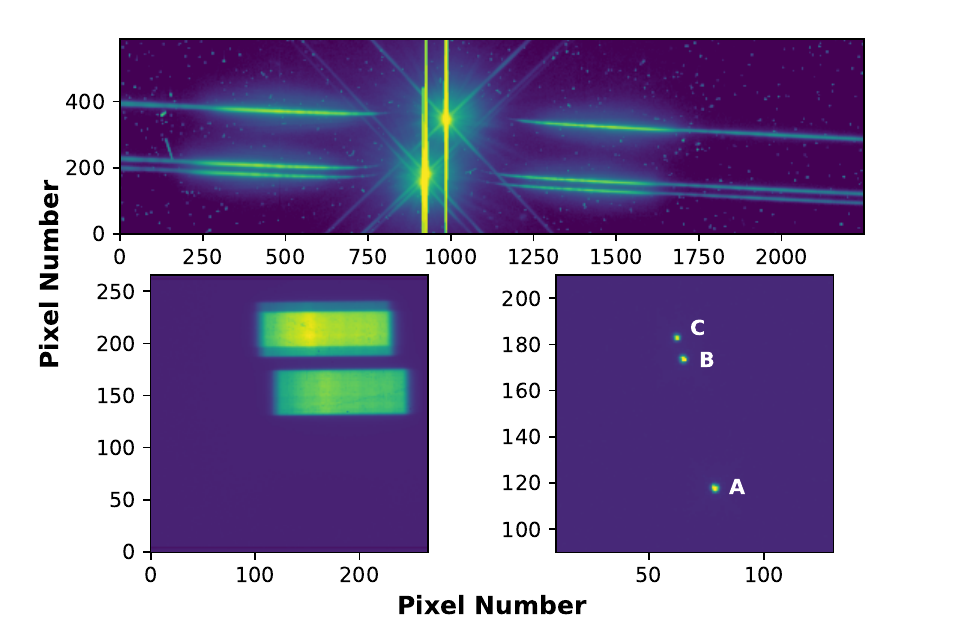}
    \caption{Examples of the initial data products. For the top image and the bottom left image, the y-direction is the cross-dispersion (spatial) direction and the x-direction is the spectral direction. \textbf{Top:} A UVIS \texttt{flt} file, where the zeroth order is imprinted in the center of the image, with the $\pm1$ orders spread to the left and right, respectively. \textbf{Bottom:} An IR \texttt{ima} file on the left and the acquisition image on the right. Color scales differ between images and are optimized to illustrate the spectra. In all images, all three stars in the system can be seen, with B and C much closer together than A. Notably, C is now the farthest star from A, instead of B as was the case with 2003 \hst/NICMOS data shown in \cite{Winters2019}. }
    \label{fig:raw_data}
\end{figure*}

\subsection{UVIS Data Reduction} \label{sec:uvis_red}

For the UVIS reduction, we begin with the \texttt{flt} files from the  Barbara A. Mikulski Archive for Space Telescopes (MAST) site\footnote{\url{https://archive.stsci.edu}}. These are fits files that have undergone standard initial calibration steps, including dark and bias subtraction, flat field correction, and initialization of the error array. 

\subsubsection{Flare Seen in LTT~1445C}

\begin{figure*}
    \centering
    \hspace*{-1cm}
    \scalebox{0.75}{\includegraphics{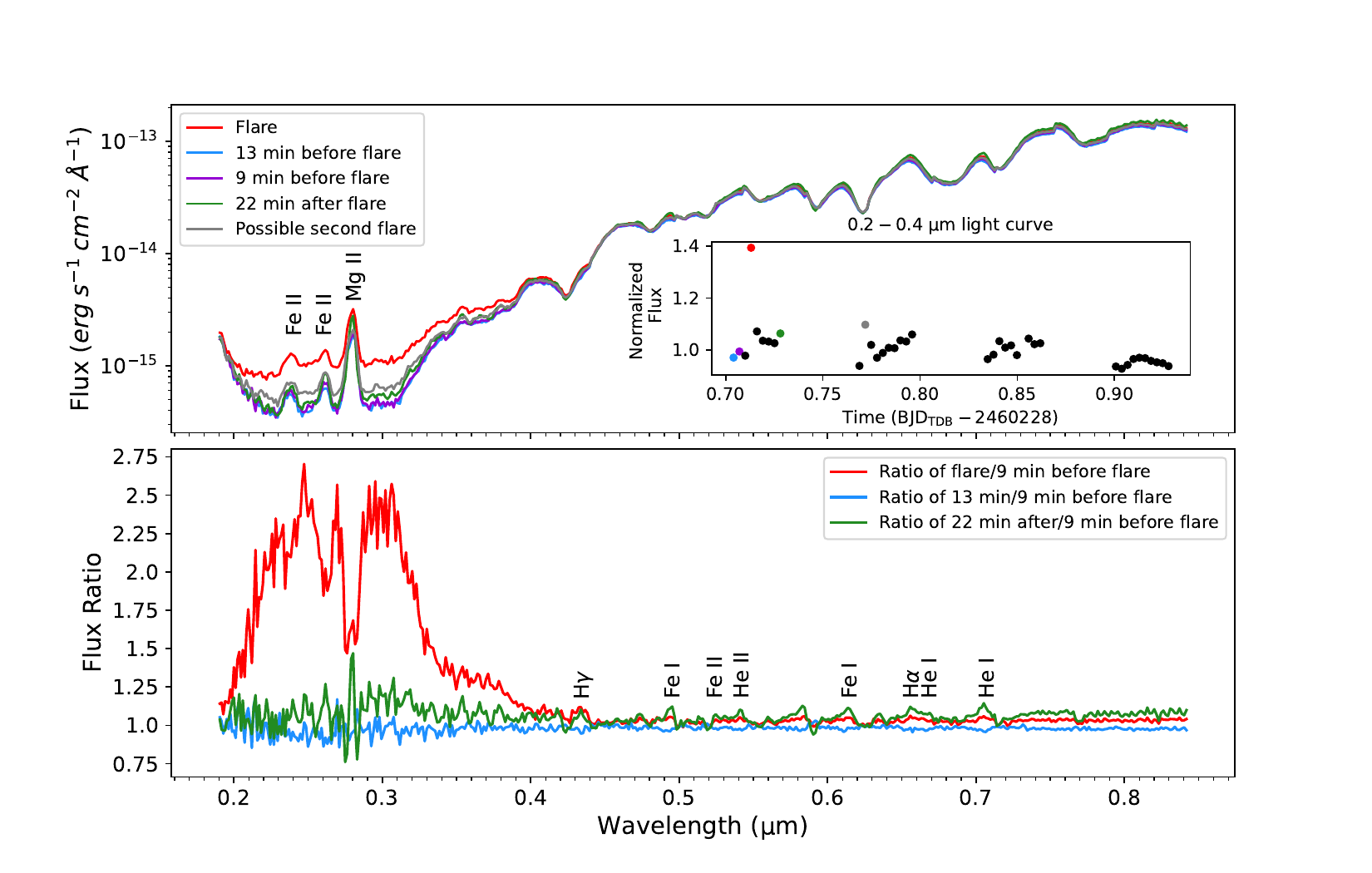}}
    \caption{\textbf{Top:} Stellar spectrum and light curve of LTT1445~C, showing a flare occurred in the fourth exposure of the third UVIS observation. The light curve is integrated from $\rm 0.2-0.4\;\mu m$, the region where flare shows flux enhancement. Colored points in the light curve match their corresponding spectra. The post-flare spectra shown (in green) also demonstrates slightly increased flux in the same region. There also appears to be a second, smaller flux in the second orbit. \textbf{Bottom:} The flux ratio between the flare and quiescent state (red), the post-flare elevated state and a quiescent state (green), and two quiescent states (blue). The flare shows a flux enhancement factor of $\sim2.25$, but the Mg II emission line does not follow this trend. There are also chromospheric optical emission lines seen in the flare and even more so in the post-flare environment.}
    \label{fig:flare}
\end{figure*}

All three stellar spectra can be seen with WFC3/G280, as seen in \autoref{fig:raw_data}. Upon examining the raw data, we discovered a flare that had occurred during the third UVIS observation (on 2023~October~11) on LTT~1445C, the smaller of the neighboring stars. 

We bring this flare to the reader's attention because only a handful of NUV spectroscopic observations for flares exist in the literature (e.g., \citealt{Hawley1991, Wargelin2017, Kowalski2019}), and, to the best of our knowledge, this is the first simultaneous and continuous NUV/optical spectra taken of a flare with a single instrument.
(\citealt{Kowalski2019, Macgregor2021}, and \citealt{Paudel2024} have conducted multiwavelength studies using multiple instruments and combinations of broadband photometry and low-resolution spectroscopy.) Additionally, there is increasing concern that flare frequencies/NUV energies based on optical surveys are greatly underestimated (e.g., \citealt{Brasseur2023, Paudel2024}), which in turn may mean we are underestimating the influence flares have on the photochemical evolution of a planet's atmosphere. This LTT~1445C flare is only apparent in the UV ($\rm <0.4\;\mu m$) and thus adds to this concern.

We follow the initial data reduction steps detailed in Section 3.1.3 below to extract the stellar spectrum for the $+1$ order of LTT~1445C. To minimize contamination from the nearby LTT~1445B star, we use a narrow aperture half-width of 3 pixels. We convert from counts ($e^{-}$) to monochromatic flux following \cite{Pirzkal2020}: we divide by the sensitivity curve for the $+1$ order (units of $e^{-}\;{\rm s^{-1}/erg\;s^{-1}\;cm^{-2}\;\AA^{-1}}$), and then divide by the exposure time (185.5 s) and the bin width per pixel to determine the flux per \AA. We confirm our result is reasonable by estimating the flux using STScI's Exposure Time Calculator (ETC)\footnote{\url{https://etc.stsci.edu}}.

We show the stellar spectra of the flare, as well as several exposures taken before and after the flare, in \autoref{fig:flare}. The inset in the top panel shows the light curve integrated from $\rm 0.2-0.4\;\mu m$, with colored exposures corresponding to the colored spectra in the top panel. The flare (shown in red) is clearly visible, with its spectrum displaying excess flux between $\rm 0.2-0.4\;\mu m$ relative to the quiescent state before the flare (shown in blue and purple). Additionally, an exposure taken after the flare (shown in green) still shows elevated activity. There also appears to be a second, smaller flare in the second orbit (shown in the grey in the inset). 

In the bottom panel of \autoref{fig:flare}, we show the ratio of the flare spectrum to a quiescent state (taken 9 minutes before the flare) in red. In blue, we show the ratio between two quiescent exposures (taken 13 and 9 minutes before the flare) and in green, the ratio between the elevated post-flare state (taken 22 minutes after the flare) and quiescent state. The flare ratio is striking, with the flare reaching over $2\times$ the flux of the quiescent state between $\rm 0.25-0.32\;\mu m$. Though the shape of the flare spectrum and ratio spectrum appears non-thermal, if the peak represents the true peak of the blackbody flux, this signifies the flare reached $\sim10,000$ K, which agrees well with past M dwarf flare studies (e.g., \citealt{Hawley2003, Kowalski2013}). It is important to note, however, that the fractional throughput of the G280 grism drops below 0.1 shortward of $\rm \sim0.21\;\mu m$, making this region of the spectrum the least reliable. If this is what causes the apparent decrease of the flare flux at the shortest wavelengths in \autoref{fig:flare}, we may not have truly captured the peak of the flare. Indeed, some recent studies are questioning the reliability of the notional $9,000-10,00$~K blackbody that is typically assumed for flares (e.g., \citealt{Howard2020, Jackman2023, Berger2024}), noting that some flares might have higher near-UV and far-UV fluxes that are not captured in optical flare surveys. Interestingly, the strong Mg II emission line seen even in the quiescent state does not show the same flux increase as the rest of the flare. We hypothesize that this could be because the flare raises the base of emission, thereby decreasing the strength of these chromospheric emission lines. 

When one examines the post-flare elevated state to the quiescent pre-flare state (green line in bottom panel), optical emission lines are seen from a variety of neutral and ionized atomic species. We use \cite{Muheki2020},
\cite{Bakowska2021}, 
and the NIST Atomic Spectra Database\footnote{\url{https://www.nist.gov/pml/atomic-spectra-database}} 
to identify these lines. We were unable to confidently identify the line at $\rm 0.583\;\mu m$, which may be V~I. Notably, these emission lines are more enhanced after the flare, as has been seen previously by \cite{Kowalski2013} and \cite{Macgregor2021}. This phenomenon has been interpreted as evidence for chromospheric evaporation \citep{Fisher1985} during flares, in which non-thermal particles in the corona accelerate downward to impact the chromosphere, heating it to coronal temperatures. This then causes chromospheric thermal emission as the chromospheric plasma expands into the corona. (This explains the so-called Neupert effect \citep{Neupert1968}, the correlation between nonthermal and thermal radiation seen during flares.)

The happenstance discovery of this flare supports the findings from \cite{Brown2022} and \cite{Rukdee2024} that LTT1445~B and C are more active than LTT1445A, and C specifically is the dominant source of X-ray emission in the system. This finding is also supported by \cite{Diamond-Lowe2023}, who found H$\alpha$ emission (a common activity indicator) in both the B and C component, but not in A. Clearly, LTT~1445C contributes to the overall high-energy environment of the entire LTT~1445 system. With B and C only 34~au from A and its planets, the activity from the neighboring stars could indeed impact the planets' evolution, as discussed in \cite{Rukdee2024}. 

\subsubsection{UVIS Diffraction Spike Correction}

We now return to our target star, LTT~1445A. 
In \autoref{fig:raw_data}, one can see that LTT~1445B and C have overlapping zeroth orders and, importantly, the diffraction spikes from these zeroth orders cross over the $+1$ order from LTT~1445A. In order to recover the blue-most end of the $+1$ order, we empirically correct for these diffraction spikes.

To do this, for each exposure we determine the shape of the diffraction spike along each row in the vicinity of the LTT 1445A $+1$ spectrum. We then calculate the median diffraction flux shape per every 40 rows, and subtract the median background flux from this median diffraction spike. Finally, per each set of 40 rows, we subtract the respective median diffraction spike. We choose to do a median per 40 rows because the diffraction spike slowly changes amplitude in the cross-dispersion direction. A median per 40 rows is the best balance between including enough rows to capture the true morphology of the diffraction spike while accounting for the changing amplitude of the feature.

\subsubsection{UVIS Initial Data Reduction} \label{sec:uvis_initial_reduction}

We use the \firefly reduction suite \citep{Rustamkulov2022, Rustamkulov2023}, which has been validated by several \jwst studies (\citetalias{Lustig-Yaeger2023} \citeyear{Lustig-Yaeger2023}, \citetalias{May2023} \citeyear{May2023}, \citetalias{Moran2023} \citeyear{Moran2023}) to reduce the data and fit the light curves. In this analysis, \firefly has been modified for use with \hst/WFC3 UVIS data. 

Starting with the \texttt{flt} files, we first remove the cosmic rays using the \texttt{lacosmic} routine \citep{vanDokkum2001}. In the second and third visits, we manually remove a cosmic ray that lay close to the spectrum using the background from neighboring pixels. Next, we measure the x-position and y-position pixel shifts as the telescope slightly moves over the course of the observation. Using these shifts, we interpolate all exposures onto a uniform x-y grid, after which we measure the trace of the spectrum. 

We extract the stellar spectrum for both the $+1$ and $-1$ orders using a box summation per column. For each visit, after testing aperture half-widths between 9--15 pixels for the $+1$ order and 6--14 pixels for the $-1$ order, we selected the aperture size that minimized the scatter in the out-of-transit white light curve data. For the $+1$ order, we use an aperture half-width of 12, 11, and 12 pixels for the first, second, and third visits respectively. For the $-1$ order, our half-widths are 10, 11, and 9 pixels, respectively. We note that changing the aperture size does have meaningful impacts on the systematics seen in the white light curves. Surprisingly, the directionality of this change is not always predictable, changing visit-to-visit. At times,  sharp ramps are seen in the $+1$ light curve, as illustrated in the UVIS white light curves (\autoref{fig:uvis_wlcs}). These ramps, as expected, decrease with increasing aperture size in the first visit, as the S/N is optimized, yet they actually increase with increasing aperture size for the second visit. In the third visit, increasing the aperture size decreases the ramps in the first two orbits while increasing the ramps in the second two orbits. 

Next, we get the calibrated wavelength solution using the location of the source in the acquisition image. We crop the edges of the stellar spectra so that the minimum and maximum wavelengths line up as closely as possible between both orders and between all three visits. However, the wavelength solution as a function of pixel number (the dispersion solution) is not quite linear and is different between the $+1$ and $-1$ orders, making it challenging to co-add the orders appropriately. To address this, we attempted to interpolate the $-1$ order onto the $+1$ order wavelength solution, but we determined that because the dispersion solutions are different, this interpolation ended up altering the scatter of the transmission spectrum. Instead, we treat the $+1$ and $-1$ orders as separate data sets through the light curve fitting. See Section 3.1.4 for details on how the two orders are summed together for the final transmission spectrum.

\subsubsection{UVIS Light Curve Fitting}

\begin{figure*}
    \centering
    \hspace*{-1.75cm}
    \scalebox{0.72}
    {\includegraphics{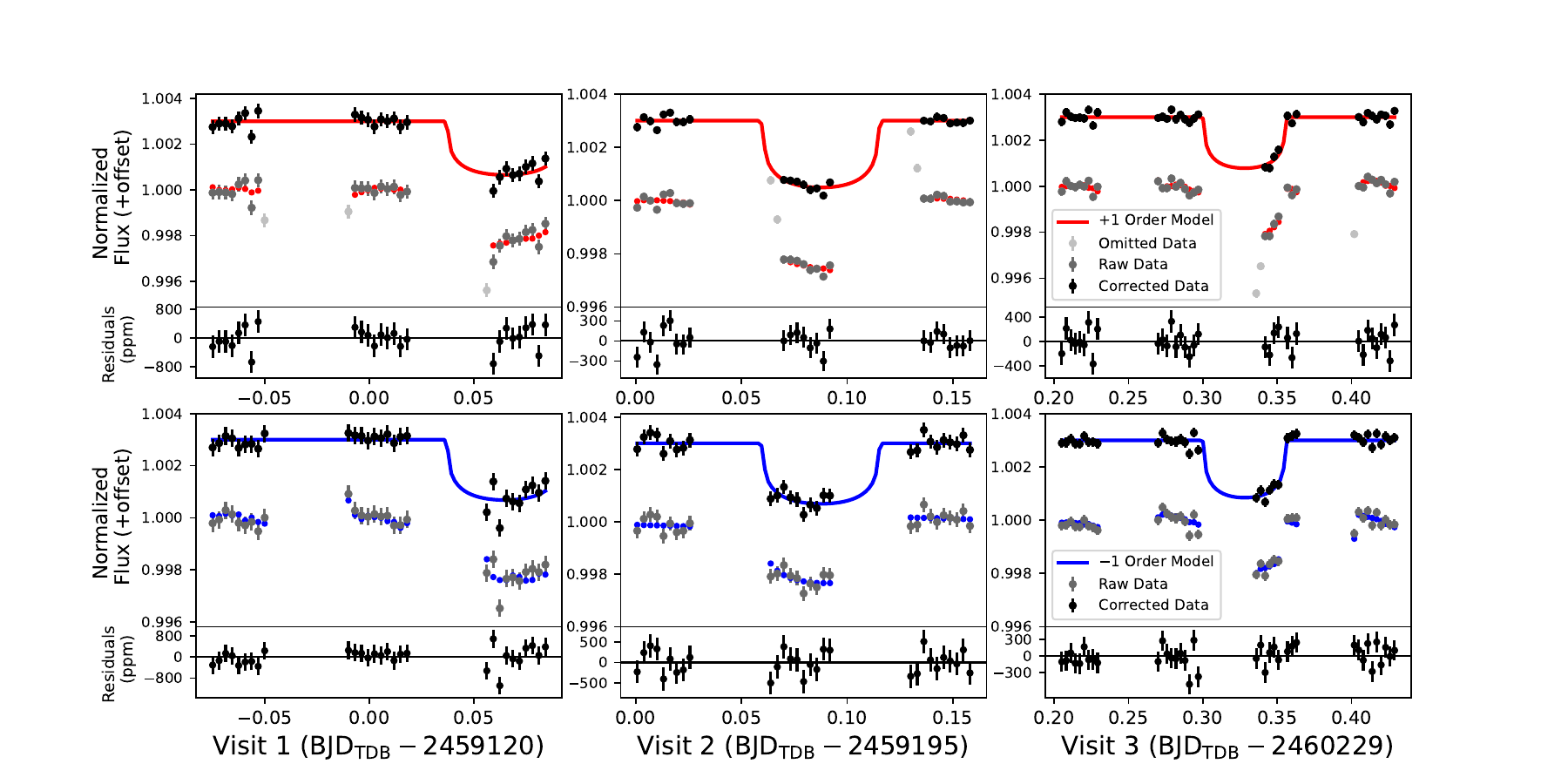}}
    \caption{UVIS (G280) white light curves for all three observed transits. The $+1$ and $-1$ orders are fitted separated and are shown on the top and bottom rows in red and blue, respectively. The raw light curve data are shown in dark grey, with the light grey points in the $+1$ order being the outlier points excluded in the white light curve fits. The model fits to the data are shown as red and blue points, and their residuals are shown individually below each plot. The full phase model, along with the systematics-removed points, are shown offset above the raw data, seen as a solid line and black points, respectively.}
    \label{fig:uvis_wlcs}
\end{figure*}

From the extracted stellar spectra, we sum data in wavelength space to produce white light curves for the $\pm1$ orders as shown in \autoref{fig:uvis_wlcs}. For the $+1$ order, we remove either the first or the first and second exposures from the final two orbits of each observation during white light curve fitting. These points 
are strong outliers and contribute to the sharp ramp effect described above in Section 3.1.3. Interestingly, though this effect was not reported in the \hst/WFC3 UVIS observations of HAT-P-41b \citep{Wakeford2020}, it has been noted with \hst's STIS instrument \citep{Brown2001, Sing2011, Sing2019}. However, with STIS, the flux of the first exposure was always meaningfully lower than the rest of the data, whereas in this case, it is lower in visits 1 and 3 but higher in visit 2. The excluded points are shown in \autoref{fig:uvis_wlcs} as light grey points, while the raw data is shown in dark grey. We test both including and excluding these points. Including them causes the white light curve to require a much more complicated systematics model, and thus much larger error bars, but does not change the shape of the transmission spectrum. We therefore argue that the best method is to exclude them in the white light curve. 

The two orders have different instrument systematics and we therefore fit both the white and spectroscopic light curves separately between the two orders, summing them together only in the final transmission spectrum. \autoref{fig:uvis_wlcs} shows all six fits to the white light curve data (three visits, each containing a $+1$ and $-1$ order fit). The raw data is shown in dark grey, with the model fits as red or blue points (for the $+1$ and $-1$ orders, respectively) overlaid on the raw data; the residuals between these are illustrated beneath each light curve. Offset from the raw data, we also visualize the full phase light curve, along with the systematics-removed data shown as black points. 

To fit the white and spectroscopic light curves, we consider that the total flux is a contribution of the transit model $T(t, \theta)$ (dependent on the time $t$ and orbital phase $\theta$), systematics model $S(x)$ (dependent on various optical state parameters $x$), and flux from the star itself:
\begin{equation}
    f(t)=T(t, \theta) \times S(x) \times F_0.
\end{equation}

To determine $S(x)$, which varies between each light curve, we follow the jitter decorrelation method introduced by \cite{Sing2019}. This entails detrending the data using a systematics model that includes optical state vectors as well as the ``traditional" detrending parameters used for \hst transit data (e.g., \citealt{Huitson2013, Wakeford2013, Nikolov2014, Demory2015, Sing2016, Lothringer2018}). The optical state vectors are extracted from the time-tagged engineering jitter files (\texttt{jit} fits files) for each science exposure (see \citealt{Sing2019} for further details), while the ``traditional" detrending parameters include the \hst orbital phase $\phi_{HST}$, the x-position and y-position shifts of the point-spread function (PSF) on the detector, and a linear slope in time $\phi_t$. The traditional detrending parameters typically account for the well-known thermal breathing of \hst \citep{Brown2001b}, while the optical state vectors, which describe the performance of various pointing controls, allow us to detrend against the pointing stability of the telescope itself. To determine which combination of optical state vectors and detrending parameters to use for each light curve, we test every combination of these factors and begin by choosing the combination that has the lowest Akaike information criterion (AIC) score. However, because many of the optical state vectors are not independent from one another, our resulting systematics model is often overly complex and included vectors that had high degrees of covariance. Indeed, \cite{Sing2019} showed that including both pairs of a correlated pair of vectors lead to larger fitting degeneracies. We therefore exclude optical state vectors with a Pearson correlation coefficient with another vector greater than 0.6. For each order and visit, we show the optimal systematics model employed in \autoref{tab:wlc_systematics}.

\begin{deluxetable}{ccc}
\tablecaption{Detrending parameters, including optical state vectors from the engineering jitter files, used in the light curve fitting for the $\pm1$ orders for all three UVIS observations. \label{tab:wlc_systematics}}
\tablenum{2}
\tablewidth{0pt}
\def\arraystretch{.85}
\tablehead{\colhead{Date of Observation} & \colhead{$+1$ Order} & \colhead{$-1$ Order}}
\startdata
    2020 September 27 & $RA$, $LimbAng$, & \textit{V2\_dom}, $\rm \phi_{HST}^{*}$\\
    & $Lat$, x-shift$^{*}$ & \\
    &  & \\
    2020 December 11 & y-shift$^{*}$, $\rm \phi_{HST}^{*}$ & $Long$, $\rm \phi_{t}^{*}$\\
    &  & \\
    2023 October 11 & \textit{V3\_roll}, $Lat$ & \textit{V3\_dom}, \textit{V2\_roll},\\
    & x-$\rm shift^{*}$, $\rm \phi_{t}^{*}$, & $RA$, $Long$, \\
    & $\rm \phi_{HST}^{*}$ & $\rm \phi_{HST}^{*}$\\
\enddata
\tablecomments{Optical state vector keyword descriptions can be found on the STScI Jitter File Format Definition webpage\footnote{\url{https://www.stsci.edu/hst/instrumentation/focus-and-pointing/pointing/jitter-file-format-definition}}. $^{*}$ Denotes a ``traditional" detrending parameter, i.e., not an optical state vector.}
\end{deluxetable}

For the transit model $T(t, \theta)$, we use \texttt{batman} \citep{Kreidberg2015}. Because of the nature of \hst's orbit around the Earth, we are not able to collect data for the entire transit. This means that in our light curve fitting, we only fit for the flux and transit depth, $(R_p/R_s)^2$. All other astrophysical parameters are kept fixed to their values from \cite{Winters2022} (shown in \autoref{tab:sys_params}): inclination ($i$), semi-major axis in units of stellar radii ($a/R_s$), and the orbital period ($P$). Mid-transit time ($T_0$) is calculated from the $T_0$ and period given in \cite{Winters2022} assuming a circular orbit. We note that \cite{Winters2022} gives only an upper limit of 0.110 for the eccentricity of \planetname. Additionally, though \planetname is in a multiplanet system and could be subject to transit-timing variations (TTVs), \cite{Winters2022} estimates that the TTVs will be on the order of only 1 minute. 

For limb darkening, we utilize the quadratic model using $q_1$ and $q_2$ as our coefficients, as propounded by \cite{Kipping2013}, where 
\begin{equation} \label{eq:_q1q2}
\begin{split}
q_1 & =(u_1+u_2)^2,\\
q_2 & =0.5u_1(u_1+u_2)^{-1}.
\end{split}
\end{equation}
\noindent We use the \cite{Hauschildt1999} stellar models and fix $q_1$ and $q_2$ (see \autoref{tab:wlc_params}) based on the wavelength range covered by WFC3/UVIS. The white light curves are fit using \texttt{emcee} \citep{emcee2013}, and best fit $(R_p/R_s)^2$ values are shown in \autoref{tab:wlc_params}.

\begin{deluxetable}{c c} 
\tablecaption{White light curve limb darkening coefficients and ${(R_p/R_s)^2}$ values. \label{tab:wlc_params}}
\tablenum{3}
\tablewidth{0pt}
\def\arraystretch{.85}
\tablehead{\colhead{Parameter} & \colhead{Value}}
\startdata
& \textbf{Fixed} \\
$q_1$ (WFC3/UVIS) & 0.576 \\
$q_2$ (WFC3/UVIS) & 0.260 \\
$u_1$ (WFC3/G141) & 0.0838 \\
$u_2$ (WFC3/G141) & 0.341 \\
\hline
\textbf{WFC3/G280 (UVIS)} & $\mathbf{(R_p/R_s)^2}$ \\
2020 September 27 $+1$ order & 
$0.00194\pm0.00012$ \\
2020 September 27 $-1$ order & 
$0.00189\pm0.00011$ \\
2020 December 11 $+1$ order & 
$0.00204\pm0.00005$ \\
2020 December 11 $-1$ order & 
$0.00188\pm0.00009$ \\
2023 October 11 $+1$ order & 
$0.00180\pm0.00011$ \\
2023 October 11 $-1$ order & 
$0.00175\pm0.00009$ \\
\textbf{WFC3/G141 (IR)} & $\mathbf{(R_p/R_s)^2}$ \\
2021 January 07 forward & $0.00192\pm0.00004$ \\ 
2021 January 07 reverse & $0.00190\pm0.00003$ \\
\enddata
\tablecomments{
Limb darkening coefficients ($u_1$, $u_2$, $q_1$, $q_2$) are calculated from stellar models from \cite{Hauschildt1999}. $a/R_s$, $i$, and $T_0$ are fixed to their literature values from \cite{Winters2022}, shown in \autoref{tab:sys_params}.} 
\vspace{-1cm}
\end{deluxetable}

We fit the spectroscopic light curves for the wavelength-dependent transit depth in the same way as the white light curve, except we fix the limb darkening per spectroscopic bin. Our error bars from the spectroscopic fits are rescaled if needed such that $\chi_{\nu}^2=1$ for each spectroscopic light curve, in order to appropriately describe the scatter of the data around our best-fit model.

\firefly utilizes a binning scheme that strives for equal counts per bin, instead of equal widths in wavelength space, which is important when building a systematics model. This is particularly true for data with low S/N, such as the blue-end of the LTT~1445A UVIS data, where the photospheric flux from the M-dwarf host drops precipitously toward zero. If we were to bin in equal wavelength space, the error bars would increase drastically at the blue end, and our systematics model would not be nearly as constrained. 

Because the dispersion solution is different between the two orders, when co-adding the binned spectra we must take care that we are co-adding pixels that correspond to the same wavelengths. To do this, we apply the standard \firefly binning procedure to the $+1$ order, determine what pixel index in the $-1$ order corresponds to the $+1$ wavelength bins, and bin up the $-1$ order based on these pixel indices. Only then do we co-add the $+1$ and $-1$ orders together.

\begin{figure}
    \centering
    \hspace*{-0.75cm}
    \scalebox{0.64}{\includegraphics{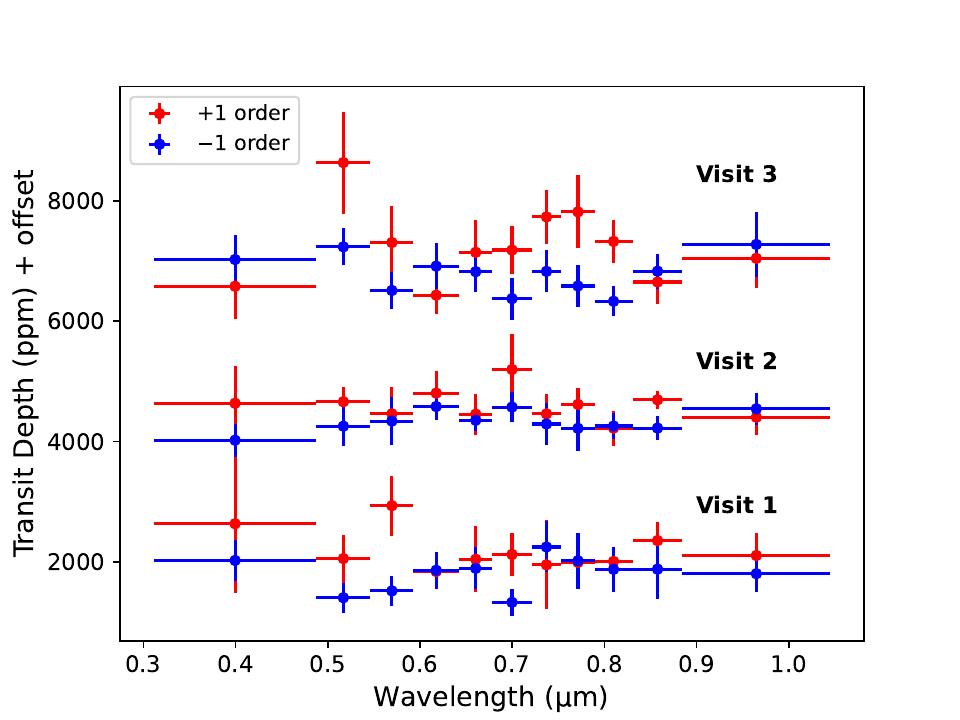}}
    \caption{UVIS transmission spectrum for all three visits, with an offset of 2500 ppm applied between each visit to visually separate the spectra. $+1$ and $-1$ orders are shown separately in red and blue, respectively, allowing a comparison within each visit. The orders agree well, with all but two points agreeing within $2\sigma$.} 
    \label{fig:uvis_spectrum_all_six}
\end{figure}

\begin{figure}
    \centering
    \hspace*{-0.75cm}
    \scalebox{0.64}{\includegraphics{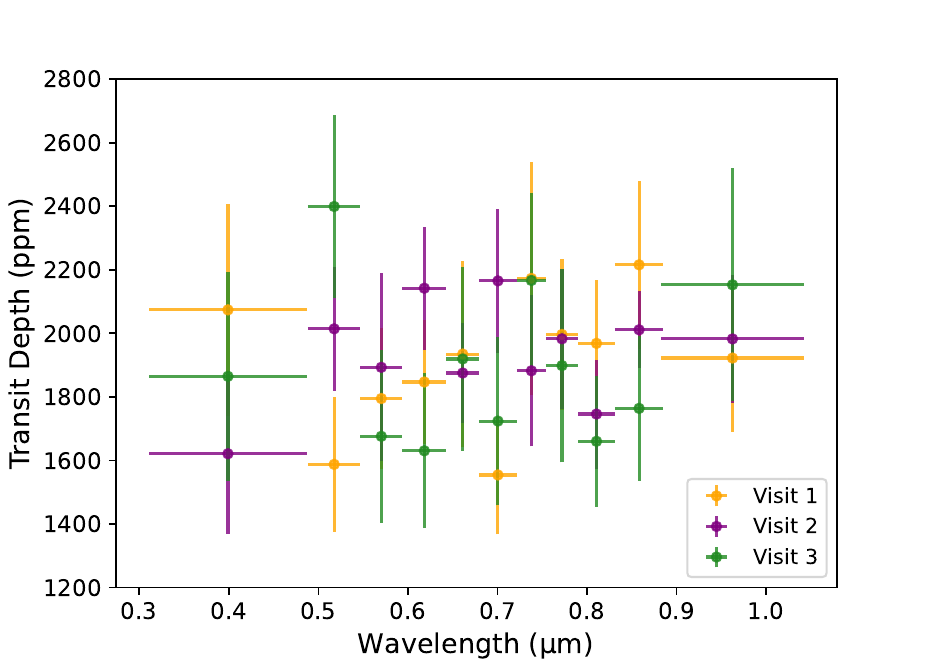}}
    \caption{UVIS transmission spectrum for all three visits, showing the weighted mean of the $+1$ and $-1$ orders for each visit. The visits agree very well.}
    \label{fig:uvis_spectrum_orders_combined}
\end{figure}

All six UVIS spectra are shown in \autoref{fig:uvis_spectrum_all_six}, with arbitrary offsets applied between each of the visits for visual clarity. We can see that overall, the $+1$ and $-1$ orders agree well, and the spectra are featureless. \autoref{fig:uvis_spectrum_orders_combined}, instead of offering an intra-visit comparison between each order, shows the inter-visit comparison, with each set of points representing the weighted mean of the $+1$ and $-1$ orders. Again, the visits show good agreement. We also apply an empirical test to determine if our error bars are accurate by comparing the scatter of all six points per wavelength bin against the calculated error bars in that bin, and we found that the two values per bin are comparable. 

The final weighted mean of all three visits is shown in \autoref{fig:spectrum_combined}, along with the WFC3/G141 (IR) reduction detailed in the following section. 

\subsection{IR Reduction} \label{sec:ir_reduction}

\subsubsection{IR Initial Data Reduction}

To reduce the G141 IR observations, we use the open-source custom pipeline \texttt{WRECS}, first described and utilized by \cite{Stevenson2014}, which was the precursor to the widely-utilized \eureka pipeline used in \jwst reductions \citep{Bell2022}. We begin our reduction with the Intermediate MultiAccum (\texttt{ima}) fits files, which have undergone almost the same calibrations steps as the \texttt{flt} files. The main difference between these files is the \texttt{ima} files includes a nonlinearity correction, as well as subtraction of the zeroth group per integration. 

As can be seen in the lower right of \autoref{fig:raw_data}, the use of the spatial scanning mode for the IR data brings the spectrum from LTT~1445A close to the overlapping spectra of LTT~1445B and C. To ensure we do not inadvertently include photons originating from B and C, we utilize ``difference images" up the ramp; in other words, in each integration, we subtract group 2 from group 1, group 3 from group 2, etc. We also compute a mask to cover the B and C spectra in the differenced images. We do this by finding the midpoint between the LTT~1445A and LTT~1445B/C spectra in the cross-dispersion direction. We then examine every pixel above this cutoff in the direction of the LTT~1445B/C spectra and mask the pixel if it exceeds a certain flux threshold. 

Next, we perform background subtraction on the differenced images, and calculate the x-position and y-position pixel shifts, applying a rough (integer-level) pixel drift correction at this stage. From here, we reject outliers from the full frame in the time direction, and finally apply a sub-pixel correction of the x- and y-position shifts. 

To extract the stellar spectrum, we perform optimal spectral extraction based on \cite{Horne1986}. This utilizes a weighted sum of pixels in the cross-dispersion direction within a given aperture width, along a measured trace. After testing aperture half-widths between 10--60 pixels, we choose an aperture half-width of 25 pixels to minimize the scatter in the white light curve. This width is sufficient to maximize spectrophotometric accuracy while minimizing noise and any residual stellar contamination from LTT~1445B and C. 

\subsubsection{IR Light Curve Fitting}

\begin{figure*}
    \centering
    \hspace*{-0.75cm}
    \scalebox{0.6}
    {\includegraphics{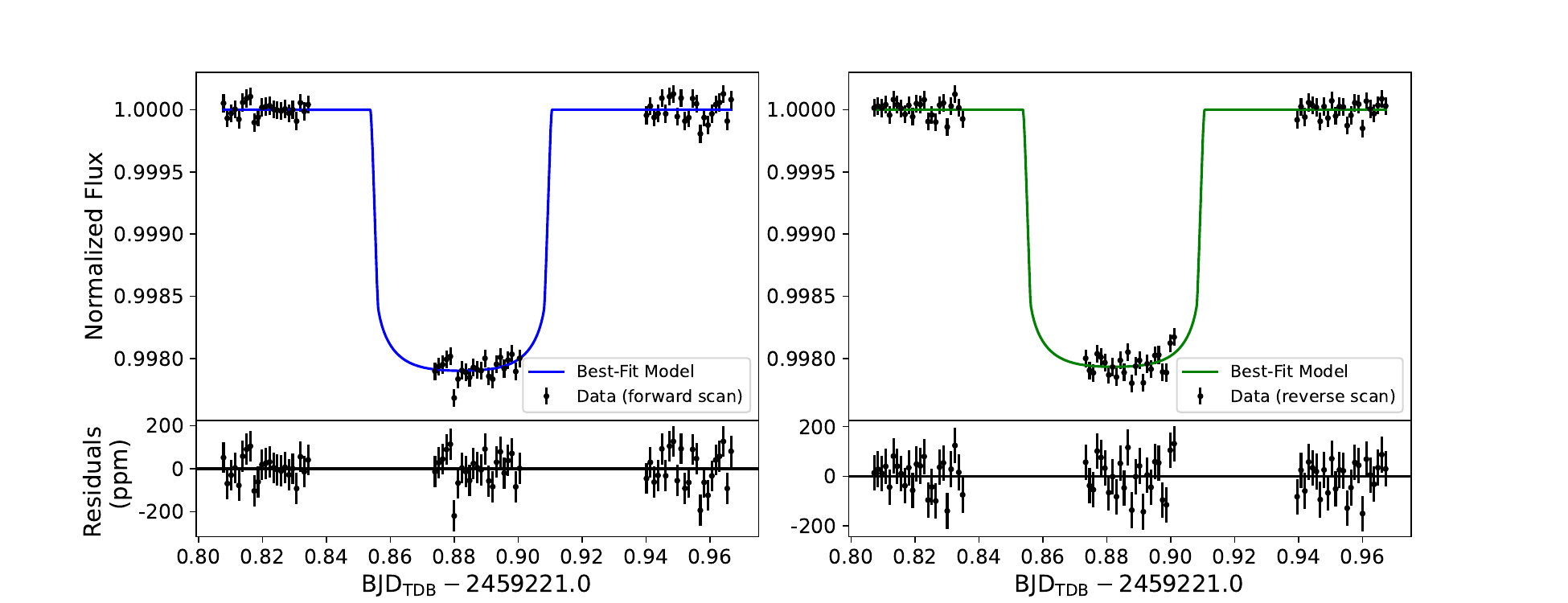}}
    \caption{The IR (G141) white light curve fits, with systematics removed, for both the forward (blue, on the left) and reverse (green, on the right) scans. The unbinned data are overplotted with the best-fit model, and the residuals for each fit are shown below. There is good agreement between the $R_p/R_s$ values calculated from each fit (see \autoref{tab:wlc_params}.) }
    \label{fig:ir_wlc}
\end{figure*}

We first model a fit to the white light curve in order to 1) calculate an up-to-date $(R_p/R_s)^2$, and 2) to apply a common-mode correction at the spectroscopic fitting stage. As in the UVIS reduction, for the light curve fitting, we fix $T_0$, $i$, $a/R_s$, and $P$ to the values listed in \autoref{tab:sys_params}, and only fit for the flux and transit depth, $(R_p/R_s)^2$. We use the quadratic limb darkening law with the classic $u_1$ and $u_2$ coefficients \citep{Kopal1950}, and we fix these coefficients (see \autoref{tab:wlc_params}) using the stellar models from \cite{Hauschildt1999}. Though this is a different parameterization than used in the UVIS reduction, because we are fixing, not fitting, limb darkening, it actually does not make a difference in the physical interpretation and error propagation of the planet radius. Using these parameters, we model the transit using the analytic approach propounded by \cite{Mandel2002}. 

To account for instrument systematics, we use a linear term and a rising exponential term following \cite{Stevenson2014} to account for the slow linear trend and ramp effect seen in each orbit. For the IR data, then, the systematics model $S(x)$ can be expressed as
\begin{equation}
    S(x) = 1 - e^{-p_1 \phi+p_2} + p_3 \phi,
\end{equation}

\noindent where $p_i$ are fitted coefficients and $\phi$ refers to the orbital phase. The linear slope ($p_3 \phi$) is particularly sharp in the first orbit, so for the white light curve, we fit a separate linear term for each orbit, so as to improve our $R_p/R_s$ fit. The second instrument systematic, the ramp effect ($1-e^{-p_1 \phi+p_2}$), is well-established in the literature (e.g., see \citealt{Berta2012, Deming2013, Swain2013, Kreidberg2014, Stevenson2014}) and is believed to be due to charge-trapping \citep{Berta2012}, similar to that reported for \spitzer data \citep{Agol2010}, in which the photodiodes retain some fraction of the electrons generated by the incident flux instead of reading out every single electron. These excess electrons are then released on a subsequent exposure as excess dark current until a steady-state is reached, creating the ``ramp effect". 

When fitting the white light curve, we fit the forward and reverse light curves independently and do not use the first orbit due to the stronger ramp and linear slope seen in this orbit. We use Differential-Evolution Markov Chain Monte Carlo (DEMCMC) model fitting \citep{terBraak2006} using the Metropolis–Hastings algorithm to find the best-fit transit model to our data. Our best-fit value for $R_p/R_s$ for each direction is shown in \autoref{tab:wlc_params}, and our white light curve fits along with their residuals are shown in \autoref{fig:ir_wlc}. We see good agreements between the forward and reverse scans. Though the linear systematic has still not been perfectly removed from the second orbit (the first set of data points in \autoref{fig:ir_wlc}, since the first orbit is removed for the white light curve fits), as seen in the correlated noise in the residuals, this is still a much better fit than can be achieved with using a single linear term for the entire observation. 

To create the transmission spectrum, we utilize the same DEMCMC technique to model each light curve over a given spectroscopic bin. To do this, we apply a common-mode correction to each spectroscopic light curve. This entails dividing each spectroscopic bin by the white light curve, which has the best-fit transit model removed, to create a non-parametric model to use in the spectroscopic fits. This means that the non-wavelength-dependent instrument systematics (including the ramp) are easily removed at this stage, which can greatly improve precision. 

\begin{figure*}
    \centering
    \hspace*{-0.5cm}
    \scalebox{0.7}
    {\includegraphics{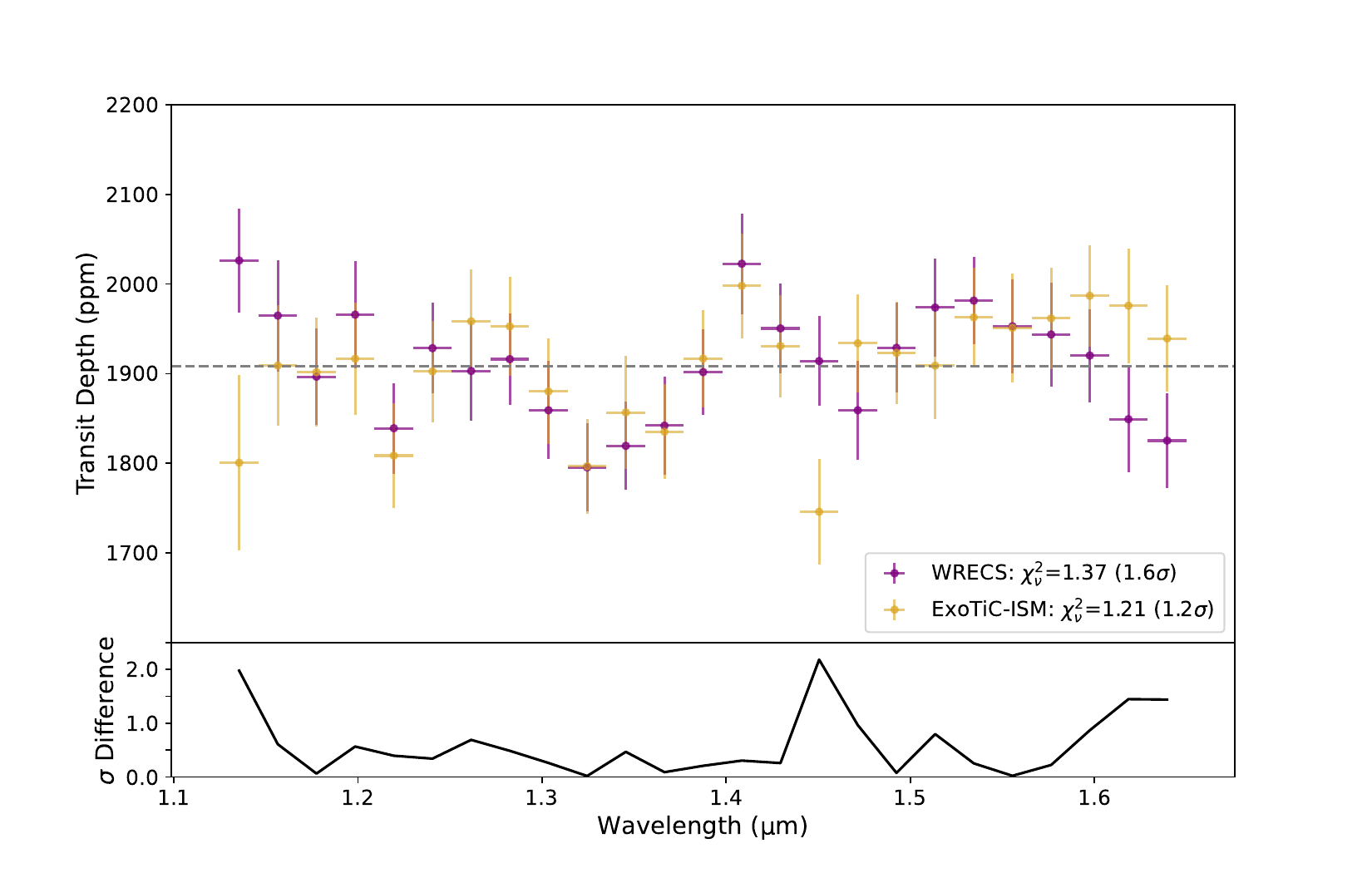}}
    \caption{\textbf{Top:} Comparison between the primary IR reduction detailed in this paper (\texttt{WRECS}), shown in purple, and the comparison \exotic reduction (offset by 148~ppm), shown in orange. A flat line, representing no atmosphere, is shown as a gray dashed line. The $\chi_{\nu}^2$ value and $\sigma$ rejection between the model and a flat line is shown in the legend for each reduction. \textbf{Bottom}: Agreement between reductions, shown as the $\sigma$ difference between the two. The reductions largely agree within 1$\sigma$.} 
    \label{fig:ir_spectrum_comparison}
\end{figure*}

At this stage, we model the forward and reverse scans together, and include all four orbits. However, we exclude the first six points from the first orbit, as the ramp in this orbit has a small wavelength dependence that is mitigated by removing these points. Because the ramp and linear slope differences per orbit are corrected by the common-mode correction, our only instrument systematic at the spectroscopic stage is a single linear term, which can have a slight wavelength-dependence. We use the same fixed astrophysical parameters and transit model as in the white light curve, and we fix $u_1$ and $u_2$ per spectroscopic bin using the stellar models from \cite{Hauschildt1999}. As in the UVIS light curve fitting, we rescale the error bars in our spectroscopic fits (if necessary) to ensure $\chi_{\nu}^2=1$ for each spectroscopic light curve. The transmission spectrum is shown in \autoref{fig:ir_spectrum_comparison}, along with a comparison reduction detailed in Section 3.2.3.

\subsubsection{Comparison with \exotic reduction} \label{sec:exotic_details}

We compare our IR reduction to another done with \exotic \citep{Wakeford2016,laginja2020}. \exotic also begins the reduction with the \texttt{ima} files output from the \texttt{CalWF3} pipeline. Like the \texttt{WRECS} reduction, we separate the data into forward and reverse scan components and treat each separately throughout the reduction. We first trim the top of the image to remove the portion of the detector containing the flux from the companion stars, resulting in a subarray of [240,176]. The stellar spectrum is extracted from the 2D spectral image using the difference imaging method where each successive non-destructive read is subtracted from each other to build up the full scan. A top-hat filter is applied to each difference image with a box aperture of 32 pixels setting all other value to zero. The reconstructed scan is then the sum of the masked difference images from which we extract the spectrum of the star. 

We then produce light curves for the forward and reverse scans separately by summing the flux across all wavelengths to produce the white light curve and in 13\,nm  bins to produce the spectroscopic light curves. We discard the first orbit of each light curve and the first exposure in each orbit as these often exhibit differing systematics due to settling of the detector. \exotic implements Instrument Systematic Marginalization, which fits the data with a grid of 50 polynomial models that account for differing observatory and instrument systematics approximating a stochastic systematic model determination process. The polynomial systematics take the form,
\begin{equation}
    S(t,\lambda) = T_1 \phi_t \times \sum^{n}_{i=1}p_i \phi_{\rm HST}^i \times \sum^{n}_{j=1}l_j \delta^{j}_{\lambda}, 
\end{equation}
where $T_i$, $p_i$, and $l_j$ are all fit parameters (set to zero if not used in the model), $\phi_t$ is the time component, $\phi_{\rm HST}$ is the HST orbital phase, and $\delta_\lambda$ is the shift in the position of the stellar spectrum with time computed from the cross-correlation of the 1D stellar spectra to a median template spectrum. Both $\phi_{\rm HST}$ and $\delta_\lambda$ are considered up to a 4th order polynomial. 
For each systematic model fit, we calculate the negative log-likelihood and convert this to a normalized weighting using the AIC. The assigned weight is then used to marginalise over all fit parameters to obtain the marginalized value \citep[see][for details]{Wakeford2016,wakeford2018}. 
We find that both forward and reverse scan measurements assign the highest weight to the same systematic model, which requires: a linear trend in time, a third order polynomial for HST phase and fourth order for $\delta_\lambda$. All light curves strongly disfavour models that do not account for a linear slope in time, resulting in those realizations contributing negligible weight to the final marginalised values.

For both our white light and systematic light curves, we fix the system values and center of transit time to the values shown in \autoref{tab:wlc_params} and fit only for $R_P/R_S$. For consistency, we additionally fix our limb-darkening parameters to those of \texttt{WRECS},
which assumes a quadratic limb-darkening law as described in Section 3.2.3. The final transmission spectrum is computed from the weighted average of the forward and reverse scan spectra. 

We show the final NIR transmission spectrum, including both reductions, in \autoref{fig:ir_spectrum_comparison}. The \texttt{WRECS} reduction is shown in purple, and the \exotic reduction is shown in orange. We have applied a 148 ppm offset to the \exotic reduction to align the weighted mean between the two visits. This offset comes early on in the reductions, as the \exotic white light curves are deeper than those of \texttt{WRECS} (i.e., it is not due to \texttt{WRECS}'s use of a common mode correction.) The mean depth of \texttt{WRECS}, not \exotic, agrees well with the mean depth found in the UVIS reduction (shown in \autoref{fig:spectrum_combined}), and so we offset \exotic to \texttt{WRECS} and not vice versa. 

From \autoref{fig:ir_spectrum_comparison}, we see that overall, the spectra agree well, with most binned data points agreeing within $1\sigma$. The bin in the \exotic reduction that appears as an outlier (at 1.45$\rm \mu m$) has been visually inspected but no clear data points or statistics of the light curve point to a culprit. We use \exotic as a check on the \texttt{WRECS} reduction, and mostly focus on the \texttt{WRECS} reduction for the remainder of the analysis presented. (We do, however, note that we double check our findings with both reductions in Section 4.4.) 

\subsection{UVIS and IR spectra together}
\begin{figure}
    \centering
    \hspace*{-0.6cm}
    \scalebox{0.63}{\includegraphics{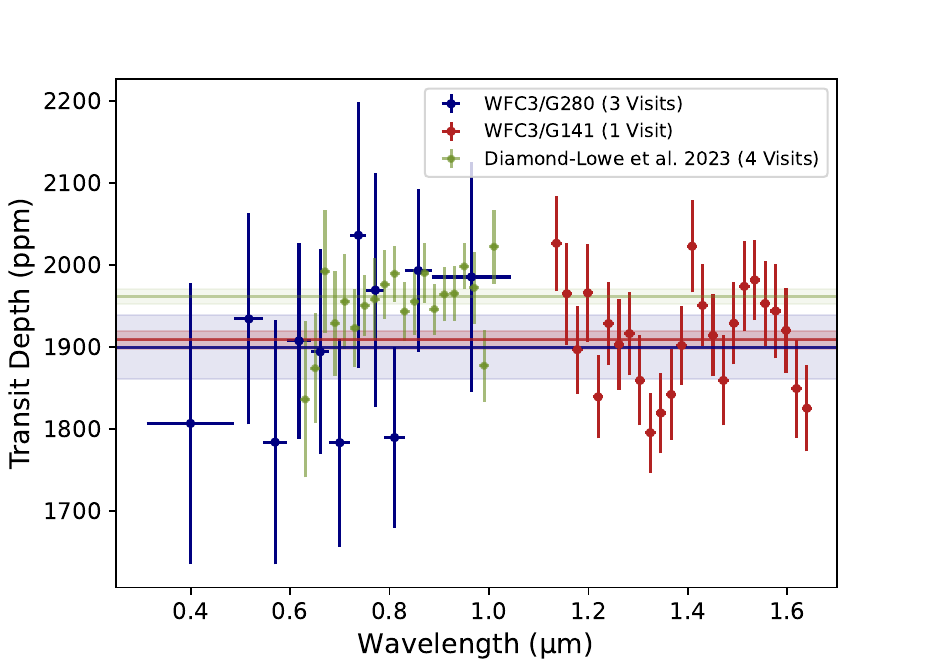}}
    \caption{Combined UVIS and IR spectrum, shown alongside the ground-based Magellan data from \cite{Diamond-Lowe2023}. Mean depths and their corresponding $1\sigma$ errors are shown as solid lines and shaded regions, respectively. The \hst  mean depths for each instrument agree well.}
    \label{fig:spectrum_combined}
\end{figure}

We plot our final spectrum, including both the UVIS and IR data, in \autoref{fig:spectrum_combined}, with the UVIS data shown in blue and the IR data shown in red. The solid lines and shaded regions represent the mean depth and $1\sigma$ error of each spectrum. Though we do not \textit{a priori} know if the two spectra ought to agree, there is no apparent offset between the two spectra, and  the mean depths show remarkable accordance, agreeing to $0.2\sigma$.  

Taken together, the UVIS and IR data are consistent with a flat line ($\chi_{\nu}^2=1.09$; rejected at 0.97$\sigma$; see \autoref{tab:statistics_comparison}). When considering the IR data alone, a flat line is a poorer fit, but still consistent with, the data: there is a 1.6$\sigma$ difference ($\chi_{\nu}^2=1.37$) between a flat line and the \texttt{WRECS} reduction and a 1.2$\sigma$ difference ($\chi_{\nu}^2=1.21$) compared to the \exotic reduction. 

In \autoref{fig:spectrum_combined}, we also compare our reduction to the optical data from \cite{Diamond-Lowe2023}, shown in green, which proves consistent with the UVIS data. Though the ground-based data is more precise than the UVIS data, it does not have as broad a wavelength coverage, and, perhaps more importantly, there is a $3.7\sigma$ offset in the mean depth between the ground-based optical and IR data. Because it is unknown whether this offset is astrophysical, telluric, or instrumental in nature, we focus on only the \hst data for the remainder of this paper.

\section{Interpretation of \planetname'\MakeLowercase{s} Transmission Spectrum} \label{sec:analysis}

Visually, potential features are evident in the IR spectrum, particularly at $\rm 1.4\;\mu m$. While we cannot rule out a flat line, in this section, we examine whether these potential features are likely due to correlated noise, stellar contamination, or hints of a real planetary atmosphere. 

\subsection{IR Correlated Noise is Negligible}
\begin{figure}
    \centering
    \hspace*{-0.5cm}
    \scalebox{0.6}
    {\includegraphics{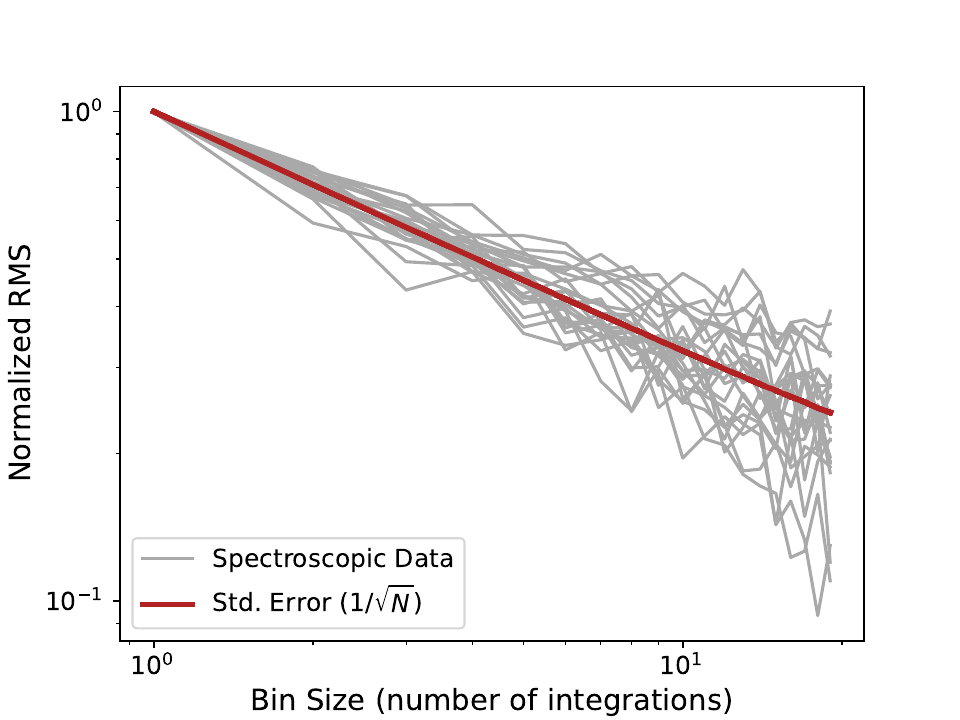}}
    \caption{Allan variance plot for the NIR spectroscopic data. This depicts (normalized) RMS scatter versus bin size, where the data are binned in time. The solid red line represents the expected decrease in scatter as $\sqrt{N}$ assuming Poisson statistics. Each grey line is a spectroscopic bin. One can see that there is negligible correlated noise.}
    \label{fig:allan_var}
\end{figure}

We examine the possibility of correlated noise in the IR spectrum through the use of an Allan variance plot. \autoref{fig:allan_var} illustrates how the RMS scatter for each spectroscopic bin decreases as one bins in time, compared to the expected decrease in scatter, which goes as $\sqrt{N}$ assuming Poisson statistics. We can see that there is negligible correlated noise, lending credence to our claim that any possible features in the spectrum are not due to unaccounted-for red noise. 

\subsection{Stellar Contamination Investigation} \label{sec:stellar_cont}
While the small size of M dwarfs makes them ideal for observing the transit signals of terrestrial exoplanets, they are also problematic. In particular, magnetically active M dwarfs can have significant inhomogeneities on their surfaces that are unocculted during transit, which imprints wavelength-dependent signatures in transmission spectra \citep[``stellar contamination", see e.g.][]{Rackham2018}. This contamination can be in the form of dark cold spots, bright hot spots, or both. (We note that the exoplanet community typically refers to these simply as ``spots" and ``faculae", following the nomenclature in the solar physics community. However, because we are not invoking a specific physical mechanism for these spots, we use more generic language here.) As this contamination effect is especially strong in the shorter wavelengths, we carefully consider whether our spectrum of \planetname can be explained by stellar contamination. This is when the UVIS data becomes absolutely critical, as it reaches shorter wavelengths than \jwst or ground-based studies. 

We approach this problem in two ways: 1) through stellar forward modeling, allowing us to place upper limits on the overall extent of variability expected, and 2) by assuming the variability is smaller than our noise limit and conducting stellar contamination retrievals to determine if the spectrum can be explained solely by contamination. 

\subsubsection{Stellar Contamination Forward Models}

We first use forward models to determine the maximum extent of possible stellar contamination due to unocculted cold or hot spots in the spectrum. We follow the method detailed by \cite{Sing2011}, in which, at a given level of photometric variability $\Delta f$ for a given wavelength, $\lambda_0$, the apparent change in transit depth due to the presence of spots is given by
\begin{equation}
    \frac{\Delta d}{d}=\pm \Delta f_{\lambda_0} \left[\left(1-\frac{F_{\lambda}^{T_{\rm spot}}}{F_{\lambda}^{T_{\rm eff}}}\right) \middle/ \left(1-\frac{F_{\lambda_0}^{T_{\rm spot}}}{F_{\lambda_0}^{T_{\rm eff}}}\right)\right],
\end{equation}

\noindent where $T_{\rm spot}$ and $T_{\rm eff}$ are the temperatures of the spots and effective temperature of the star, respectively, and $F_{\lambda}$ is the monochromatic flux of the spot/photosphere. We assume that these spots are time-averaged and always present on the stellar disk. The last term on the right-hand side (in the denominator) is a normalization factor, as the change in depth must be normalized to the bandpass in which the change in flux was measured or estimated. Because of this normalization factor, $\frac{\Delta d}{d}$ must be multiplied by $\pm1$ for cold vs. hot spots, respectively. 

$F_{\lambda}$ is calculated using the library of PHOENIX stellar models provided by France Allard and collaborators \citep{Allard2003, Allard2007, Allard2012} and available online through STScI\footnote{\url{https://archive.stsci.edu/hlsps/reference-atlases/cdbs/grid/phoenix/}}. We assume $\rm {log}$$\;g=5.0$ and [M/H]$=0.0$. We assess the impact of cold spots between 2600--3000~K and of hot spots between 3600--4400~K. We also test different magnitudes of photospheric variability $\Delta f_{\lambda_0}$. This is another important unknown for this system. \cite{Winters2019} found that spots are likely for all three stars in the system, and the variability for the entire stellar system (including all three stars) is on the order of 2\%. The authors note that much of that may be coming from the more active B and C components of the system, but interestingly enough, \cite{Kar2024} recently found the flux variability of the A component to be similar in magnitude to that of the BC component ($\sim29$ mmag or $\sim2.6\%$). 
Yet, in \cite{Winters2022}, the authors use \textit{TESS} data (after applying a dilution correction to remove the effects of B and C) to estimate that A shows only between $0.1-0.3$\% variability. 

\begin{figure*}
    \centering
    \hspace*{-1.1cm}
    \scalebox{0.78}{\includegraphics{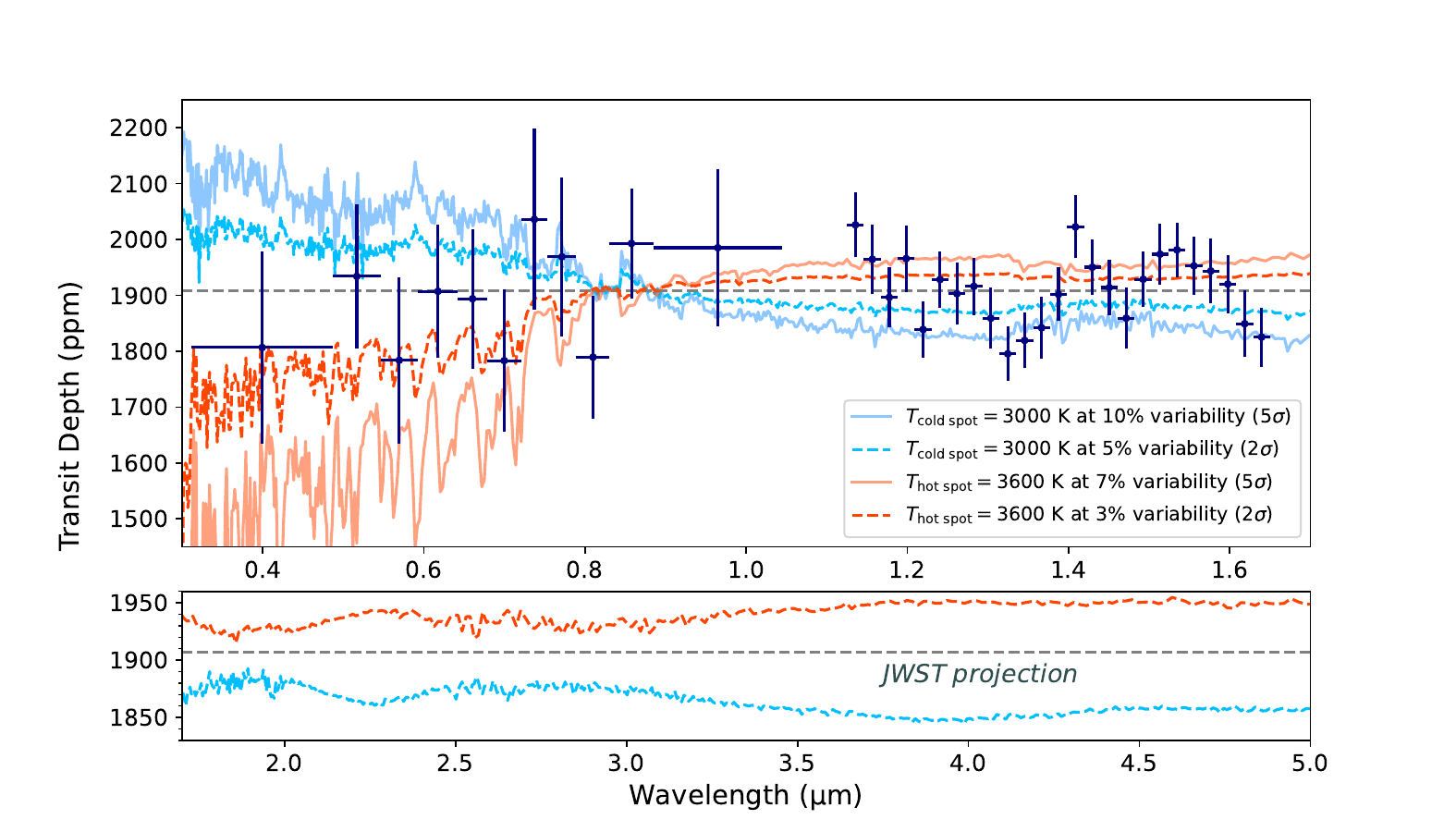}}
    \caption{Stellar contamination forward models based on \cite{Sing2011}, assuming different spot temperatures and dimming fraction (variability) for time-averaged spots that are always present on the stellar disk. \textbf{Top:} Blue lines represent a 3000~K cold spot scenario, with the solid line representing the level of variability that can be rejected to $5\sigma$ and the dashed line the level of variability rejected to $2\sigma$. Red lines are the analogous version for a 3600~K hot spot scenario. \textbf{Bottom:} Models rejected at $2\sigma$ are shown extrapolated out to \jwst wavelengths, demonstrating that the change in transit depth due to cold/hot spot contamination that can be expected is on the order of $30-40$ ppm. Mean transit depth is shown as a grey dashed line.}
    \label{fig:stellar_cont_fwd_models}
\end{figure*}

In this case, we would like to test a broad range of $\Delta f_{\lambda_0}$ values to determine what level of photometric variability can robustly be ruled out, as variability of M dwarfs remains an open question (e.g., \citealt{Mignon2023}). We test between 0.1--10\% variability, using $\rm \lambda_0=0.8\;\mu m$, as the only information about variability we have comes from \textit{TESS} data (which covers $\rm 0.6-1.0\;\mu m$). A subset of our results are shown in \autoref{fig:stellar_cont_fwd_models}, which shows contamination from 3000~K cold spots in blue and 3600~K hot spots in red. We find that overall, the temperature of the spots makes much less of a difference than the level of variability. For each case, we show the two levels of variability that can be ruled out at $5\sigma$ and $2\sigma$. For the cold spot scenario, this is 10\% and 5\%, respectively, and for the hot spot scenario, this is 7\% and 3\%, respectively. For the models rejected at 2$\sigma$, we extrapolate our findings out to \jwst wavelengths to place upper limits on the extent of stellar contamination expected. 

In \autoref{fig:stellar_cont_fwd_models}, one can see that, intuitively, the larger variability causes a steeper slope, with cold spots causing a negative slope in the UV/optical and hot spots causing a positive slope. The canonical stellar contamination water feature is seen in the cold spot scenario (blue lines) at $\rm 1.4\;\mu m$, though it is too low to fit the IR data. To determine what kind of stellar contamination features may be present in \jwst data, we extrapolate the weakly rejected ($2\sigma$) models to \jwst wavelengths. Though of course the precise nature of stellar contamination is always changing, this exercise gives insight into the order of magnitude that could be expected. In the lower panel of \autoref{fig:stellar_cont_fwd_models}, we see broad water features around 1.75 and 2.75  $\mu m$ in the cold spot case. Overall, the magnitude of variability is on the order of 30--40 ppm, and this is particularly apparent shortward of $\rm 3.75\;\mu m$. We advocate that the upcoming \jwst NIRSpec observations of \planetname (GO Program 2512; P.I. N. Batalha) consider these projections when analyzing the \jwst transmission spectrum for this target.

\subsubsection{Stellar Contamination Retrievals}
To further investigate the possibility of stellar contamination, we carry out retrievals with \texttt{exoretrievals} \citep{Espinoza_2019}. Based on the framework from \cite{Rackham2018}, \texttt{exoretrievals} allows us to model the spectrum created by unocculted or occulted hot and/or cold spots on the stellar surface by replacing a fraction of the photosphere with a stellar model of another temperature. We let the cold spots vary from 2300--3240~K, and the hot spots vary from 3440--5000~K, for a photospheric temperature of $3340\pm 150$ K, and let the spot size cover up to the entirety of the unocculted photosphere. We do not allow for the presence of occulted spots, as we do not see any evidence for spot crossings in our light curves. We also consider both default PHOENIX models and BT-SETTL models given the relatively cool temperature of LTT~1445A, but see no significant difference in the results. We consider models with only cold spots, only hot spots, and both components. 

\begin{figure}
    \centering
    \hspace*{-0.6cm}
    \scalebox{0.63}{\includegraphics{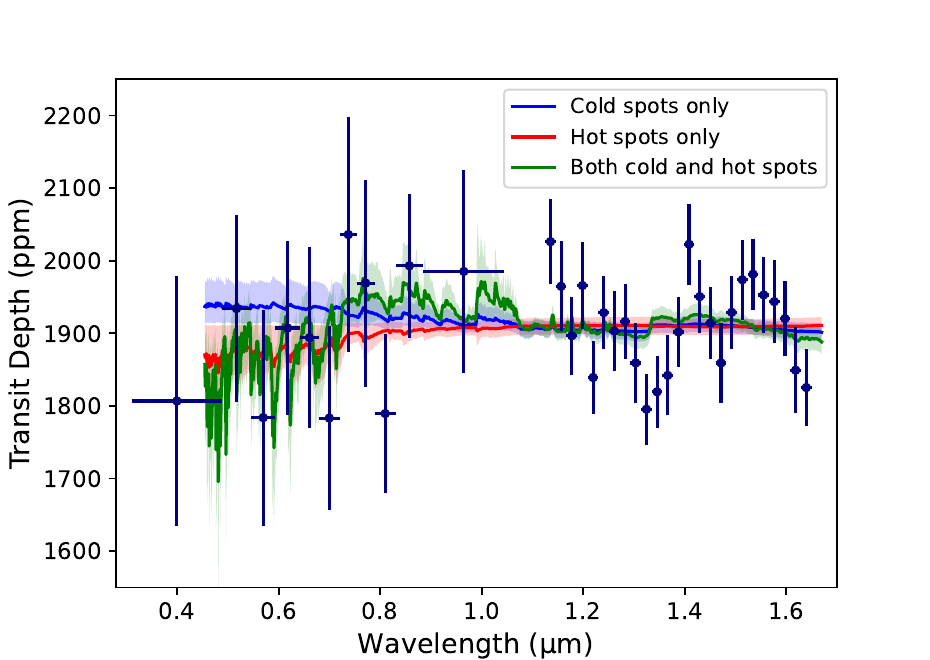}}
    \caption{Best fit stellar retrieval models using \texttt{exoretrievals} assuming the spectrum is due entirely to stellar contamination. The case of cold spots only, hot spots only, and both cold and hot spots are shown in blue, red, and green, respectively. While all three scenarios are consistent with the UVIS data, none of them are well fit to the IR data.}
    \label{fig:stellar_retrieval_NA}
\end{figure}

Even with this wide explored parameter space, none of the models are able to completely fit the feature seen in the WFC3/IR data, as seen in \autoref{fig:stellar_retrieval_NA}. Our retrieved cold and hot spot temperatures and coverage fractions are shown in \autoref{tab:stellar_retrievals_params}. 

\begin{deluxetable}{ccc}
\tablecaption{Retrieved cold/hot temperatures and coverage fraction for different model scenarios using \texttt{exoretrievals}. \label{tab:stellar_retrievals_params}}
\tablenum{4}
\tablewidth{0pt}
\def\arraystretch{.85}
\tablehead{\colhead{} & \colhead{Temperature (K)} & \colhead{Coverage Fraction (\%)}}
\startdata
        Cold Spots & $2784\pm310$ & $8\pm10$ \\
        Hot Spots & $3729\pm330$ & $16\pm22$ \\
        Cold+Hot Spots & $2898\pm250$ (cold) & $31\pm23$ (cold)\\
        & $3772\pm300$ (hot) &$26\pm25$ (hot) \\
\enddata
\end{deluxetable}

To be completely thorough, since the IR and the final UVIS observation were taken a significant period apart from each other ($\sim$ 3 years, much longer than the stellar rotation period of 85 days) and therefore may be representative of different stellar surfaces and/or levels of activity, we also consider the WFC3/IR data alone. Again, in this test we are not able to fit the IR feature well. Therefore, we continue on to the consideration of atmospheric features.

\subsection{Forward Modeling Constrains LTT~1445Ab's Atmospheric Metallicity}

If indeed the features we see are real, their size can help place constraints on the expected metallicity of the planetary atmosphere. We use \texttt{TRIDENT}, the transmission spectra forward modeling arm of the \texttt{POSEIDON} atmospheric retrieval framework \citep{MacDonald2017, MacDonald2023} to generate models of possible transmission spectra of \planetname assuming thermochemical equilibrium. \texttt{TRIDENT} produces model atmospheres by using the \texttt{FastChem} equilibrium chemistry grid \citep{Stock2018, Stock2022}, which provides molecular and atomic mixing ratios of a myriad of species at a given pressure and temperature. By inputting stellar and planetary parameters (stellar and planetary radius, temperature, and surface gravity) as well as atmospheric metallicity and C/O ratio, one can generate mixing ratios of specified species as a function of pressure (uniform in log space with 100 layers) using the \texttt{FastChem} grid. Then, opacity sampling from \texttt{POSEIDON}'s opacity database \citep{MacDonald2022} is used at a specified spectral resolution to generate a model spectrum. 

For \planetname, we use planet and stellar parameters as defined in \autoref{tab:sys_params}. We use $T_{\rm eq}=424$~K, set $\rm C/O=0.55$ (roughly solar) and explore metallicities between 10--1000$\times$ solar. Though the C/O ratio can have a meaningful impact on planetary spectra, particularly when and if $\rm C/O>1$, (e.g., \citealt{Madhu2012, Moses2013}), low $\rm C/O$ ratios may be more likely in M-dwarf systems \citep{Tadashi2016}. What's more, within the \texttt{TRIDENT} framework, changing the C/O ratio from subsolar to solar (we tested between 0.2 and 0.55) only impacts the resulting forward model by about 10 ppm, which is well within our error bars. We include as trace species in the forward models the dominant species $\rm H_2O$, $\rm CO_2$, $\rm CO$, $\rm N_2$, $\rm CH_4$, $\rm NH_3$, and $\rm HCN$, as these species have features within the WFC3/G141 bandpass or may be important contributors to the overall atmospheric chemistry.

\begin{figure}
    \hspace*{-0.6cm}
    \centering
    \scalebox{0.63}{\includegraphics{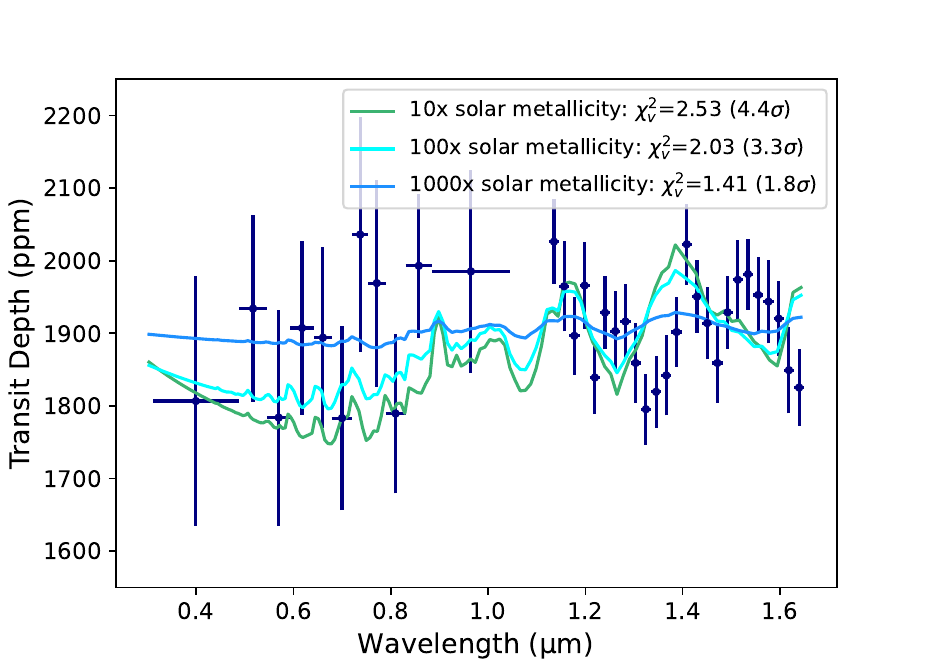}}
    \caption{10$\times$, 100$\times$, and 1000$\times$ solar metallicity forward models generated by \texttt{TRIDENT} assuming thermochemical equilibrium and $\rm C/O=0.55$. Each forward model is uniquely vertically shifted to find the minimum possible $\chi^2_{\nu}$. This $\chi^2_{\nu}$ value, along with the corresponding $\sigma$ rejection, is reported in the legend. 1000$\times$ solar metallicity is favored over lower metallicity atmospheres.}
    \label{fig:fwd_modeling}
\end{figure}

Results from this forward model test are shown in \autoref{fig:fwd_modeling}, with the 10$\times$, 100$\times$, and 1000$\times$ times solar metallicity cases shown in green, cyan, and blue, respectively, against the backdrop of the transmission spectrum. \texttt{FastChem} computes mixing ratios assuming thermochemical equilibrium and, in all cases, $\rm H_2O$ becomes the dominant trace species with features in the G141 bandpass. Thus, the forward models are dominated by $\rm H_2O$ opacity (clearly seen just shy of 1.2 $\rm \mu m$ and at 1.4 $\rm \mu m$), with only smaller, subtle effects from the other molecules included in the model. 

Because the precise atmospheric composition is different between the different forward models illustrated, the scale heights, and thus the effective planetary radius and transit depth, are also different. To account for this, we vertically shift each model individually to the mean depth where the $\chi^2_{\nu}$ value is minimized between the data and model. This allows us to compare the $\chi^2_{\nu}$ values between models without worrying that our comparisons are confounded by offsets. The shifted spectra, and the resulting $\chi^2_{\nu}$ values are shown in \autoref{fig:fwd_modeling}. 

We also report the $\sigma$ rejection between the data and the model, with the 10$\times$, 100$\times$, and 1000$\times$ scenarios being rejected by 4.4, 3.3, and 1.8$\sigma$, respectively. This indicates that we can reject an atmosphere with substantial $\rm H_2/He$ (10$\times$ and 100$\times$ metallicities). As Neptune has a metallicity of $\rm \sim100\times$ solar \citep{Karkoschka2011, wakeford2018}, this supports the findings from \cite{Diamond-Lowe2023} that \planetname indeed does not host a primary, $\rm H_2/He$-dominated atmosphere, and it therefore is not likely a sub-Neptune in nature. The data are consistent with a 1000$\times$ metallicity atmosphere, again supporting the assumption that any atmosphere around \planetname is likely secondary in nature. 

\subsection{Atmospheric Retrievals Explore
Possible Atmospheres for LTT~1445Ab}

To fully investigate the possibility of a planetary atmosphere, we use \POSEIDON for atmospheric retrievals. \POSEIDON uses the Bayesian nested sampling package \texttt{PyMultiNest} \citep{Feroz2009, Buchner2014}  to explore the parameter space. 

We conduct several exploratory retrievals in the sections below. In each case, our free parameters are the atmospheric temperature (for which we use a prior of $\mathcal{U}$(300, 900)\,K), the planetary radius at 10 bar pressure ($\mathcal{U}$(0.85$\rm R_P$, 1.15$\rm R_P$)), the $\log_{10}$ volume mixing ratios of atmospheric gases considered (either with a uniform prior of $\mathcal{U}$(-12, -1) or with a centered-log ratio prior, as discussed below), and the $\log_{10}$ surface pressure in bar ($\mathcal{U}$(-7, 2)). Each retrieval used 2,000 MultiNest live points. The spectral models are calculated at $R=$~4,000 between $\rm 0.3-1.8\;\mu m$, with the model subsequently convolved with the instrument point-spread function (PSF) and binned to the resolution of the data. 

\subsubsection{Is An Atmosphere Favored?} \label{sec:retrieval_atmosphere_or_not_test}

We begin this exploration by assessing whether a flat line or an atmosphere is favored in a Bayesian model comparison. We would like to be agnostic about the type of atmosphere, and so we test both uniform and centered-log ratio (CLR) priors on the mixing ratios of the possible atmospheric constituents. Uniform priors assign a specific gas to be the non-parameterized filler gas (with a calculated mixing ratio of $\rm 1-\sum_i X_i$, where $\rm \sum_i X_i$ is the sum of the mixing ratios of the trace species). This places a strong prior on this background gas to be the most abundant in the atmosphere \citep{Benneke2012}. While this strategy may be appropriate when dealing with gas giants (where we can assume $\rm H_2/He$ atmospheres from bulk density measurements) it may be less so when examining rocky planets, where the atmospheric make-up is entirely unknown \citep{Benneke2012}. Instead, it may be more appropriate to test CLR priors, which allow any atmospheric species to become the bulk background gas. For more details on the mathematical description of CLR priors, see \citet{Benneke2012}. 

We therefore test both uniform priors, assuming a bulk $\rm H_2/He$ (primary) atmosphere, as well as CLR priors that allow for a variety of heavier secondary atmospheres. When using Bayesian model comparison to test for an atmosphere, one should use a relatively simple atmospheric model. We thus test several combinations of molecular atmospheric constituents, including $\rm H_2O$, $\rm NH_3$, $\rm CH_4$, and HCN, all of which have features around $\rm 1.4\;\mu m$ that could explain the potential features seen in LTT~1445Ab's spectrum. 

We show a subset of these tests in \autoref{tab:retrievals_flatline_test}, where various atmospheric scenarios are shown alongside their Bayes factor relative to a flat line and the $\sigma$ significance of detection. As can be seen, in all scenarios, the Bayes factors are small, and an atmosphere is not significantly favored over a flat line scenario. However, the choice of mixing ratio prior (i.e., whether a primary or secondary atmosphere is assumed) clearly impacts whether an atmosphere is weakly favored or not. If a primary atmosphere is assumed, a flat line is the favored explanation. On the other hand, the evidence for an atmosphere is slightly stronger (though still statistically insignificant) if a secondary atmosphere is allowed. This corroborates our evidence from the forward models that a bulk $\rm H_2/He$ atmosphere is not favored. We also ran these tests with the \exotic reduction, which found very similar results. Likewise, if we only examine the IR data (excluding the UVIS data), an atmosphere is still only weakly ($<3\sigma$) preferred. 

\begin{deluxetable}{cccc}
\tablecaption{Bayesian model comparisons for the presence of an atmosphere from \POSEIDON, showing the Bayes factor (BF) relative to a flat line and $\sigma$ detection significance. \label{tab:retrievals_flatline_test}}
\tablenum{5}
\tablewidth{0pt}
\def\arraystretch{.85}
\tablehead{\colhead{Atmosphere} & \colhead{Molecules} & \colhead{BF/$\sigma$} & \colhead{Preferred?}}
\startdata
       Primary &  & & \\
       (uniform prior) & HCN$+\rm N_2$ & 1.0/$-$ &  Flat line\\ 
       & HCN$+\rm H_2O$ & 0.46/$-$ & Flat line\\ 
       & HCN, $\rm H_2O$, $\rm NH_3$ & 0.63/$-$ & Flat line\\
       Secondary & & & \\
       (CLR prior) & HCN$+\rm N_2$ & 3.2/2.1 &  Atmosphere\\ 
       & HCN$+\rm H_2O$ & 2.5/2.0 & $-$\\ 
       & HCN, $\rm H_2O$, $\rm NH_3$ & 1.3/$-$ & $-$\\
\enddata
\end{deluxetable}

\subsubsection{What Kind of Atmosphere Could Be Present?}

To explore possible culprits of the ``features" seen in the spectrum, we conduct a Bayesian nested model comparison test, in which the Bayesian evidence between a nested and reference model is used to assess the statistical significance of a given molecule. It is important to recognize that because a flat line is not ruled out, this is merely an exploratory exercise to test which molecules are more likely than others to explain the potential features in our spectra. Because of that, we do not include every possible molecule that may be in \planetname's atmosphere, but limit ourselves to those molecules with features around $\rm 1.4\;\mu m$. Therefore our reference model includes HCN, $\rm NH_3$, $\rm H_2O$, and $\rm CH_4$, and we assume CLR priors (allowing for a heavier, presumably secondary atmosphere). We also include $\rm N_2$, $\rm H_2$, and $\rm He$ in our reference model as possible spectrally inactive ``filler" gases (also with a CLR priors).

The results of this test are shown in \autoref{tab:retrievals_comparison}. We show the Bayes factor and $\sigma$ detection significance relative to the reference model for several scenarios. In each case, we exclude one molecule from the reference model and compare the Bayesian evidence between the reference model and the new model with one excluded molecule. The Bayes factor and $\sigma$ significance shown in the table refers to the reference model relative to the new model. For example, in \autoref{tab:retrievals_comparison}, in the case in which HCN is excluded, the Bayes factor is 1.97 with a 1.8~$\sigma$ significance, demonstrating that the reference model (which includes HCN) is weakly preferred over the new model (which excludes HCN). From this test, we can see that HCN and is weakly preferred, while $\rm N_2$, $\rm NH_3$, $\rm H_2O$, and $\rm CH_4$ are not preferred at all. We also compare the reference model to a no-atmosphere scenario, but the two scenarios have almost equal Bayesian evidences and cannot be distinguished. This makes sense, as our reference model is now more complex than the models considered in Section 4.4.1. 

\begin{deluxetable}{ccc}
\tablecaption{Bayesian model comparisons for specific absorbers, showing the Bayes factor relative to a reference atmospheric model and the $\sigma$ detection significance of each molecule. \label{tab:retrievals_comparison}}
\tablenum{6}
\tablewidth{0pt}
\def\arraystretch{.85}
\tablehead{\colhead{Retrieval} & \colhead{Bayes Factor} & \colhead{$\sigma$}}
\startdata
      Atmosphere? & 1.02 & 1.0\\
      HCN & 1.97 & 1.8 \\
      $\rm N_2$ & 1.09 & No $\rm N_2$ preferred \\
      $\rm NH_3$ & 0.806 & No $\rm NH_3$ preferred\\
      $\rm H_2O$ & 0.881 & No $\rm H_2O$ preferred\\
      $\rm CH_4$ & 0.725 & No $\rm CH_4$ preferred\\
\enddata
\tablecomments{Reference model uses CLR priors with HCN, $\rm N_2$, $\rm NH_3$, $\rm H_2O$, and $\rm CH_4$ as atmospheric constituents. This comparison is used as an exploration to determine possible culprits behind the ``features" seen at $\rm 1.4\;\mu m$ and $\rm 1.55\;\mu m$.}
\end{deluxetable}

\begin{figure*}%
\hspace*{0cm}
    \centering
    \hspace*{-0.5cm}
    \includegraphics[scale=0.375]{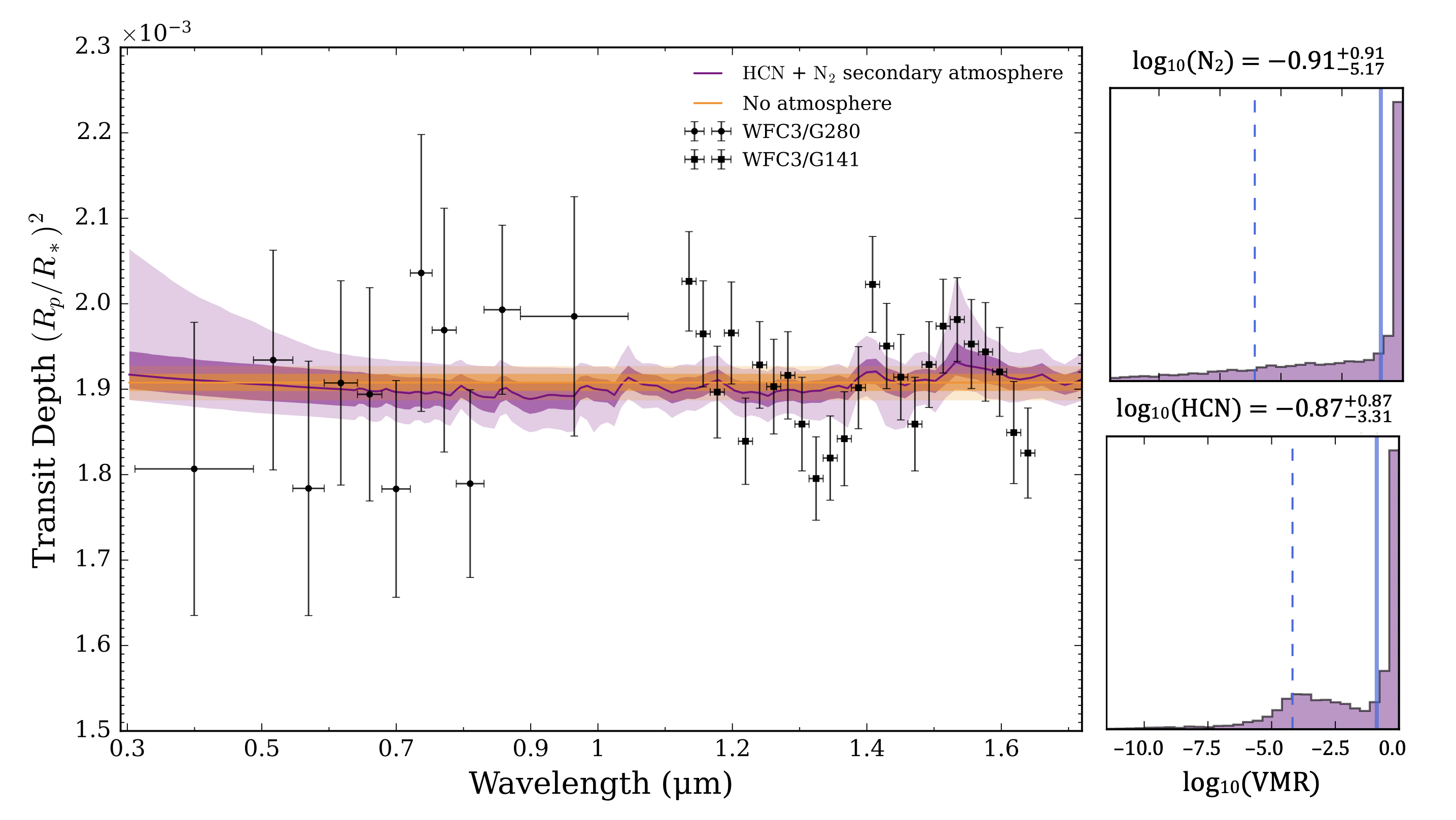}
    \caption{Best-fit retrieval on combined UVIS and IR data using \POSEIDON. CLR priors are used to allow for a high mean-molecular weight atmosphere. In purple is the atmosphere retrieval, compared to a flat line. 1$\sigma$ and 2$\sigma$ confidence on the retrieval is shown in dark and light shaded regions. Histogram from the retrieved abundances of $\rm N_2$ and $\rm HCN$ are shown on the right. $\rm HCN$ is weakly favored but not detected at statistical significance.}
    \label{fig:retrieval}
\end{figure*}

Based on our result from Tables \ref{tab:retrievals_flatline_test} and \ref{tab:retrievals_comparison}, we show the best-fit retrieval (an atmosphere with $\rm N_2$, $\rm H_2$, $\rm He$, and $\rm HCN$, assuming CLR priors) in \autoref{fig:retrieval} compared to a retrieved flat line. The atmosphere scenario is shown in purple, with 1$\sigma$ and 2$\sigma$ confidence ranges shown in the dark and light shaded regions, respectively. $\rm HCN$ accounts for the features seen at $\rm 1.4\;\mu m$ and $\rm 1.55\;\mu m$.

Retrieved abundances for $\rm N_2$ and $\rm HCN$ are shown on the right-hand side of \autoref{fig:retrieval}. When using CLR priors, high abundances are explored (corresponding to an ``uptick" in the prior when plotted in log mixing ratio space), which is indeed what is seen for both molecules. The HCN posterior displays two ``bumps" corresponding to either: 1) HCN could be a trace gas in an N$_2$-rich atmosphere ($\log$ HCN $\sim-4$), or 2) HCN could be the background gas. While the latter scenario is statistically explored, the former is more physically realistic as N$_2$ will always more readily be the dominant nitrogen-bearing molecule compared to HCN. We note that HCN is also weakly favored in the case of uniform priors (which assumes a bulk $\rm H_2$/He atmosphere), with the retrieved abundance equivalent to the ``bump'' seen at $\rm log\;HCN\sim-4$.


\section{Discussion} \label{sec:discussion}

\subsection{Comparison of All Possibilities}

In order to uniformly compare the possibility of a flat line, stellar contamination, and an atmosphere, we use the $\sigma$ rejection, $\chi^2_{\nu}$ and Bayesian Information Criterion (BIC) values to compare models calculated across different frameworks. We show our results in \autoref{tab:statistics_comparison}, which reports these values (along with the number of free parameters in each case, $k$) for a flat line, the three retrieved stellar contamination models (assuming cold spots, hot spots, and both), and the best-fit retrieved atmosphere model (allowing for a secondary atmosphere with $\rm HCN$). 

It is clear that all models are consistent with the data, as all models are within $3\sigma$ of the data. It may be tempting to conclude that an atmosphere is preferred over stellar contamination (particularly for the case of both cold and hot spots) due its lower $\chi^2_{\nu}$ and BIC. However, from considering the number of free parameters in each case, we can see that in reality, the difference in BIC values is largely driven by differences in model complexity. For the cases of only cold or only hot spots, the BIC values are nearly identical to the atmosphere case, with the latter having only a slightly lower $\chi^2_{\nu}$. 

From this, we must conclude that our data remain consistent with a flat line, and it remains difficult to distinguish between an atmosphere and stellar contamination in the assumption that the spectrum is not flat. Though visually the stellar retrievals do not fit the IR features well (see \autoref{fig:stellar_retrieval_NA}), overall the stellar retrieved models cannot be rejected. To disentangle these scenarios will require further follow-up at a broader wavelength range.

\begin{deluxetable}{ccccc}
\tablecaption{$\sigma$ rejection, $\chi^2_{\nu}$, and BIC values for various interpretations of the spectrum, with the number of free parameters ($k$) given for each scenario. $N=36$ in all cases. \label{tab:statistics_comparison}}
\tablenum{7}
\tablewidth{0pt}
\def\arraystretch{.85}
\tablehead{\colhead{Scenario} & \colhead{$\sigma$} & \colhead{$\chi^2_{\nu}$} & \colhead{BIC} & \colhead{$k$}}
\startdata
     Flat line & 0.97 & 1.09 & 4.67 & 1\\
     Atmosphere & 0.90 & 1.07 & 15.40 & 4\\
     Cold Spots & 1.25 & 1.19 & 15.52 & 4\\
     Hot Spots  & 1.23 & 1.18 & 15.52 & 4\\
     Cold+Hot Spots & 1.14 & 1.16 & 22.66 & 6\\
\enddata
\tablecomments{Retrieved atmosphere uses CLR priors with HCN, $\rm N_2$, $\rm H_2$, and $\rm He$ as atmospheric constituents.}
\end{deluxetable}

\subsection{What Is the Nature of \planetname?}

\cite{Brown2022}, \cite{Diamond-Lowe2024}, and \cite{Rukdee2024} all recently examined the high-energy spectra of all three stars in the LTT~1445 triplet system. \cite{Rukdee2024} do not detect flares from LTT~1445A using \textit{Chandra}'s Advanced CCD Imaging Spectrometer (ASIS-S) instrument. Likewise, we do not detect any flares from LTT~1445A across all three UVIS observations. However, \cite{Brown2022} do detect X-ray flares in separate observations with \textit{Chandra}/ASIS-S, and \cite{Diamond-Lowe2024} detect a flare in the UV with \hst's Cosmic Origins Spectrograph (COS). Taken together, \cite{Brown2022} and \cite{Diamond-Lowe2024} caution that flare frequencies and X-ray variability could be higher than expected for slowly-rotating M dwarfs such as LTT~1445A. On the other hand, \cite{Diamond-Lowe2024} still paints a more optimistic picture about atmospheric retention, demonstrating that if \planetname retained an initial $\rm CO_2$ budget at least one-tenth that of the Earth, it could maintain a 10 bar surface pressure. 

It does, therefore, at least seem plausible that \planetname could have retained its atmosphere, though of course this remains highly dependent on a myriad of factors, including how the spectral energy distribution (SED) of LTT~1445A changes over time, the flare frequency and energy distribution, the impact of flares from LTT1445B and C, as well as the planet's initial volatile content and the presence of a planetary magnetic field. Based on the planet's published mass and radius \citep{Winters2019, Winters2022, Lavie2023, Oddo2023}, a primary $\rm H_2/He$-dominated atmosphere for \planetname was unlikely, and we have confirmed that in this study. 

Whether or not the planet hosts a secondary atmosphere, however, remains to be seen. If the features in the WFC3/G141 spectrum are real, they hint there is HCN in the planet's atmosphere. HCN is intriguing from an astrobiological perspective, as it is one of the building blocks necessary to create amino acids, and could be present in nitrogen-rich atmospheres, particularly if $\rm C/O>1$ \citep{Chameides1981, Rimmer2019}. HCN can be formed in an atmosphere through photochemistry \citep{Zahnle1986, Tian2011}, a plausible mechanism for a planet only 0.04~au from its star, as well as through impact events or electrical discharge, i.e., lightning \citep{Chyba1992}. Interestingly, an enhancement of EUV stellar radiation relative to far-UV leads to an increase in the planetary HCN mixing ratio \citep{Rimmer2019}. It is ironic, then, that the high-energy environment of M dwarfs that may lead to atmospheric stripping is the same environment that may be favorable to biologically-important precursor molecules. In the case of \planetname, however, it is best to tread conservatively, as previous claims of HCN using WFC3/G141 have been refuted for 55~Cancri~e \citep{Deibert2021} and GJ~1132b (\citealt{Mugnai2021, Libby-Roberts2022}, \citetalias{May2023} \citeyear{May2023}). 

It is also important to note that for close-in, rocky planets like \planetname, we do not \textit{a priori} expect to find secondary atmospheres rich in $\rm CH_4$ and/or HCN. Instead, most evolutionary models predict $\rm CO_2$ and $\rm O2$-rich atmospheres \citep{Elkins2008, Gao2015, Luger2015, Dorne2018}. Interactions between an early atmosphere and magma ocean can produce a thick $\rm CO_2$ atmosphere \citep{Elkins2008}, and assuming a lack of plate tectonics to sequester $\rm CO_2$ internally, $\rm CO_2$ outgassing is predicted to be most efficient for planets in \planetname's mass range (between $\rm 2-3\;M_{\oplus}$; \citealt{Dorne2018}). Additionally, if the planet undergoes a runaway greenhouse effect, in which all the planet's water boils off into the atmosphere and is subsequently dissociated, with $\rm H_2$ lost to space (similar to Venus's fate), there is predicted to be a buildup of abiotic $\rm O_2$ and $\rm O_3$ in the atmosphere \citep{Gao2015, Luger2015}.

An atmosphere rich in $\rm CO_2$ and/or $\rm O_2$ would host a very high mean molecular atmosphere with feature sizes on the order of $10-20$ ppm - undetectable with \hst. The potential ``features" we observe are larger than this - on the order of 100 ppm - which suggests that they are either not real or, if they are real, that \planetname hosts a lower mean molecular weight atmosphere than might \textit{a priori} be expected for close-in M-dwarf rocky planets. Nevertheless, the composition of exoplanetary secondary atmosphere remains an open question (e.g., \citealt{Wordsworth2022}). More observations are required to disentangle these two possibilities.

Stellar contamination also remains a concern for this system, though we rule out variability $>5$\% due to cold spots and $>3\%$ due to hot spots with the UVIS data. We caution, however, that upcoming \jwst transmission studies of this planet be weary of drawing conclusions based solely on water, as stellar contamination can easily mimic a planetary water feature, as seen in the bottom panel of \autoref{fig:stellar_cont_fwd_models} and discussed by \citetalias{Moran2023} \citeyear{Moran2023}.

\section{Conclusion} \label{sec:conclusion}

In this paper, we have reduced and analyzed the UV/optical/NIR transmission spectrum of the nearby M dwarf rocky planet \planetname with \hst. Only 6.9 pc away, \planetname is relatively large (1.31 $\rm R_{\oplus}$) and cool (424 K), making it one of our most optimal targets for atmospheric studies of rocky M-dwarf planets. 

We reduce and analyze three transits with the WFC3/G280 (UVIS) grism and one transit with the WFC3/G141 (IR) grism. We find a flare on LTT~1445C in the third UVIS observation, and we demonstrate that this flare likely peaks at $\sim10,000$ K and shows optical chromospheric emission lines in the post-flare environment. 

While the planet spectrum is consistent with a flat line, there are potential features seen in the IR spectrum that are not likely due to correlated noise or stellar contamination. With only one IR transit, it remains to be seen whether these features are real or simply unlucky noise instances. 

We find a primary atmosphere for \planetname is unlikely, as we rule out atmospheres $100\times$ solar metallicity and less. Through atmospheric retrievals, we find in certain cases a secondary atmosphere is weakly ($\sim2\sigma$) preferred over a flat line scenario, with HCN being the molecule most likely to explain the potential features seen in the IR. 

Caution is warranted, however, as the spectrum is also consistent with stellar contamination. Though the IR features themselves are not well fit by stellar retrievals, retrieved models assuming stellar contamination are still broadly consistent ($1.1-1.3\sigma$) with the entire spectrum. Additionally, based on forward models that examine the slope of the UVIS data, variability up to 5\% (for cold spots) and 3\% (for hot spots) cannot be ruled out. This will be very important to keep in mind for the upcoming \jwst NIRSpec/G395H observations of this planet (GO Program 2512; PI N. Batalha).

Though the NIRSpec transit should help illuminate the nature of this planet, more transits at  wavelengths $\rm <2\;\mu m$ are needed to determine if the features seen in this study are real. Unfortunately, cost-saving measures are being considered for \hst operations as part of an Operational Paradigm Change Review (OPCR), which may include de-commissioning WFC3/IR modes in future HST cycles. At $K_{\rm mag}=6.5$, \planetname is actually too bright for many of \jwst's instruments that probe $\rm <2\;\mu m$, including NIRSpec/PRISM, NIRSpec/G140H, and NIRIS/SOSS. The best observing strategy with \jwst moving forward will be the NIRCam short-wavelength Dispersed Hartmann Sensor (DHS) mode that is newly available for science operations starting in Cycle 4. Still, we recommend WFC3/IR modes continue to be supported as they continue to do valuable science. 

\planetname is likely to continue to be one of the best rocky targets we have for transmission spectroscopy. To confidently determine whether or not this planet has an atmosphere will require an ensemble of efforts, including in-depth transmission spectroscopy and emission photometry with \jwst and continued monitoring of the high-energy environment of the host star. Given the possible hints of an atmosphere seen with \hst/WFC3, this planet should remain a top priority for the exoplanet community. Robustly detecting atmospheres (or confidently ruling them out) on a variety of rocky exoplanets will bring us into a new age of exoplanet characterization and bring us one step closer to understanding planetary habitability across our Galaxy.

\section*{Acknowledgments}
This work is based on observations made with the NASA/ ESA Hubble Space Telescope obtained at the Space Telescope Science Institute, which is operated by the Association of Universities for Research in Astronomy, Inc. Support for this work was provided by NASA through grants under the HST-GO-16039 program from the STScI.
H.R.W. was funded by UK Research and Innovation (UKRI) under the UK government’s Horizon Europe funding guarantee for an ERC Starter Grant [grant number EP/Y006313/1]. N.H.A. acknowledges support by the National Science Foundation Graduate Research Fellowship under Grant No. DGE1746891. N.J.M. acknowledges support from a UKRI Future Leaders Fellowship [Grant MR/T040866/1], a Science and Technology Facilities Council (STFC)
Consolidated Grant [ST/R000395/1], and the Leverhulme Trust through a research project grant [RPG-2020-82].
The authors thank J. Ih, M. Nixon, E. Kempton, S. Moran, and S. Peacock for helpful discussions. 

\vspace{5mm}
\facilities{This \textit{HST} data was taken from the Mikulski Archive for Space Telescopes (MAST) at the Space Telescope Science Institute.}

\software{\\ astropy \citep{astropy:2013, astropy:2018, astropy:2022}, \\ 
\texttt{batman} \citep{Kreidberg2015}, \\
\texttt{emcee} \citep{emcee2013}, \\
\texttt{exoretrievals} \citep{Espinoza_2019}, \\
\exotic \citep{Wakeford2016, laginja2020}, \\
\texttt{FastChem} \citep{Stock2018, Stock2022}, \\
\firefly \citep{Rustamkulov2022, Rustamkulov2023}, \\
\texttt{lacosmic} \citep{vanDokkum2001}, \\
matplotlib \citep{Hunter2007}, \\
numpy \citep{Harris2020}, \\
pandas \citep{McKinney2010}, \\
\texttt{POSEIDON} \citep{MacDonald2017, MacDonald2023}, \\
\texttt{PyMultiNest} \citep{Feroz2009, Buchner2014}, \\
scipy \citep{Virtanen2020}, \\
\texttt{WRECS} \citep{Stevenson2014}}

\section*{Data Availability} 
All of the data presented in this article were obtained from the Mikulski Archive for Space Telescopes (MAST) at the Space Telescope Science Institute. The specific observations analyzed can be accessed via \dataset[doi:10.17909/asja-zn95]{https://doi.org/10.17909/asja-zn95}.\\
The data products for Figures \ref{fig:flare}, \ref{fig:uvis_wlcs}, \ref{fig:uvis_spectrum_all_six}, \ref{fig:ir_wlc}, \ref{fig:ir_spectrum_comparison}, and \ref{fig:spectrum_combined} are available here: \url{https://zenodo.org/doi/10.5281/zenodo.13227946}.

\bibliography{main_revised_arxiv}{}
\bibliographystyle{aasjournal}

\end{document}